\documentclass[journal,onecolumn,12pt]{IEEEtran}
\ifCLASSINFOpdf
\else
\fi
\usepackage[cmex10]{amsmath}
\usepackage{theorem}
\usepackage{amssymb}
\usepackage{algorithmic}
\usepackage{algorithm}
\usepackage{graphicx}

\newcommand{\mathset}[1]{\left\{#1\right\}}

\newcommand{\bfa}{{\boldsymbol a}}

\newcommand{\bfg}{{\boldsymbol g}}

\newcommand{\bfm}{{\boldsymbol m}}

\newcommand{\bfs}{{\boldsymbol s}}

\newcommand{\bfu}{{\boldsymbol u}}
\newcommand{\bfv}{{\boldsymbol v}}

\newcommand{\bfx}{{\boldsymbol x}}
\newcommand{\bfy}{{\boldsymbol y}}

\newcommand{\bfX}{{\boldsymbol X}}

\newcommand{\cC}{\mathcal{C}}

\newcommand{\cH}{\mathcal{H}}

\newcommand{\cY}{\mathcal{Y}}

\theoremstyle{plain}
\theorembodyfont{\normalfont\slshape}

\newtheorem{thm}[algorithm]{Theorem$\!$}
\newenvironment{theorem}
{\begin{thm}\hspace*{-1ex}{\bf.}}{\end{thm}}

\newtheorem{lem}[algorithm]{Lemma$\!$}
\newenvironment{lemma}{\begin{lem}\hspace*{-1ex}{\bf.}}{\end{lem}}

\newtheorem{prop}[algorithm]{Proposition$\!$}
\newenvironment{proposition}{\begin{prop}\hspace*{-1ex}{\bf.}}{\end{prop}}

\newtheorem{cor}[algorithm]{Corollary$\!$}
\newenvironment{corollary}{\begin{cor}\hspace*{-1ex}{\bf.}}{\end{cor}}

\newtheorem{defn}[algorithm]{Definition$\!$}
\newenvironment{definition}{\begin{defn}\hspace*{-1ex}{\bf.}}{\end{defn}}

\newtheorem{xmpl}[algorithm]{Example$\!$}
\newenvironment{example}{\begin{xmpl}\hspace*{-1ex}{\bf.}}{\end{xmpl}}

\newtheorem{cnstr}[algorithm]{Construction$\!$}
\newenvironment{construction}{\begin{cnstr}\hspace*{-1ex}{\bf.}}{\end{cnstr}}

\newtheorem{asmp}[algorithm]{Assumption$\!$}

\begin{document}
\title{Rank-Modulation Rewrite Coding \\for Flash Memories}
\author{Eyal En Gad,
        Eitan Yaakobi,~\IEEEmembership{Member,~IEEE,}
				Anxiao (Andrew) Jiang,~\IEEEmembership{\\Senior Member,~IEEE,}
        and
				~Jehoshua Bruck,~\IEEEmembership{Fellow,~IEEE}% <-this % stops a space
\thanks{The material in this paper was presented in part at the
IEEE Int. Symp. on Inform. Theory (ISIT), Saint Petersburg, Russia, August 2011~\cite{EngJiaBru11}, at the IEEE Int. Symp. on Inform. Theory (ISIT), Cambridge, MA, U.S.A., July 2012~\cite{EngJiaBru12}, and at the IEEE Int. Symp. on Inform. Theory (ISIT), Istanbul, Turkey, July 2013~\cite{EngYaaJiaBru13}.
}
\thanks{
E. En Gad, E. Yaakobi and J. Bruck are with the Department
of Electrical Engineering, California Institute of Technology, Pasadena,
CA, 91125 USA (e-mail: eengad@caltech.edu; yaakobi@caltech.edu; bruck@caltech.edu).}% <-this % stops a space
\thanks{A. Jiang is with the Department of Computer Science and Engineering and the Department of Electrical and Computer Engineering, Texas A\&M University, College Station, TX 77843 
USA (email: ajiang@cse.tamu.edu).}% <-this % stops a space
}
\maketitle

\begin{abstract}
The current flash memory technology focuses on the cost minimization of its static storage capacity. However, the resulting approach supports a relatively small number of program-erase cycles. This technology is effective for consumer devices (e.g., smartphones and cameras) where the number of program-erase cycles is small. However, it is not economical for enterprise storage systems that require a large number of lifetime writes.

The proposed approach in this paper for alleviating this problem consists of the efficient integration of two key ideas: (i) improving reliability and endurance by representing the information using relative values via the rank modulation scheme and (ii) increasing the overall (lifetime) capacity of the flash
device via rewriting codes, namely, performing multiple writes per cell before erasure.

This paper presents a new coding scheme that combines rank-modulation with rewriting. The key benefits of the new scheme include: (i) the ability to store close to $2$ bits per cell on each write with minimal impact on the lifetime of the memory, and (ii) efficient encoding and decoding algorithms that make use of capacity-achieving write-once-memory (WOM) codes that were proposed recently.
\end{abstract}


\IEEEpeerreviewmaketitle

\section{Introduction}

\IEEEPARstart{R}{ank} modulation is a data-representation scheme which was recently proposed for non-volatile storage devices such flash memories~\cite{JiaMatSchBru09}. Flash memories are composed of cells which store electric charge, where the amount of charge is quantized to represent information.
%Flash memories are composed of cells, each stores a limited amount of electrical charge, which represent a small amount of information bits (typically 1 to 3 bits).
Flash memory cells are quantized typically into 2, 4 or 8 disecrate levels, and represent, respectively, 1, 2 or 3 information bits. %The quantization is performed by a comparison of a cell's charge level with a set of threshold levels.

In the proposed rank-modulation scheme, a set of $n$ memory cells represents information according to the \emph{ranking} of the cell levels. For example, we can use a set of 3 cells, labeled from 1 to 3, such that each cell has a distinct charge level. We then rank the cells according to their charge levels, and obtain one of $3!=6$ possible permutations over the set $\mathset{1,2,3}$. A possible ranking would be, for example, cell 3 with the highest level, then cell 1 and then cell 2 with the lowest level.
%, as shown in Figure \ref{fig:rewriting}. 
Each ranking can represent a distinct information message, and so the 3 cells in this example store together $\log_2 6$ bits. It is suggested in~\cite{JiaMatSchBru09} that rank modulation could significantly improve the reliability and writing speed of flash memories.

%The charge leakage is tolerated well with permutations, since it behaves similarly in spatially close cells, and thus the order of the cell levels is not likely to change. In comparison, when the information is represented by a quantization of absolute values, the leakage is more likely to introduce errors. This phenomenon was demonstrated in phase-change memories, a different type of non-volatile memories with a similar leakage property \cite{PapPozMitCloBreLamEle11}.

An important concept in rank modulation is that of \emph{rewriting}. Rewriting refers to the writing of information into the flash cells by solely increasing the cell levels (without decreasing the level of any cell). It is motivated by the fact that decreasing the cell levels is an expensive operation in flash memory, called ``block erasure''. When a user wishes to update the data stored in the memory, she increases the cells' charge levels such that they form a ranking that corresponds to the desired updated data message. The cells, however, have an upper limit on their possible charge levels. Therefore, after a certain number of updates, the user must resort to the expensive erasure operation in order to continue updating the memory. The concept of rewriting codes was proposed in order to control the trade-off between the number of data updates and the amount of data stored in each update. A similar notion of rewriting codes is also studied in conventional data-representation scheme (i.e. vectors of independent symbols as opposed to rankings), with models such as ``write-once memory''~\cite{RivSha82,YaaKaySieVarWol12}, ``floating codes'' and ``buffer codes'' (both in~\cite{JiaBohBru10}).

Rank-modulation rewriting codes were proposed in~\cite[Section IV]{JiaMatSchBru09}, with respect to a rewriting method called ``push-to-the-top''. In this rewriting method, the charge level of a single cell is pushed up to be higher than that of any other cell in the ranking. In other words, a push-to-the-top operation changes the rank of a single cell to be the highest. 
A rewriting operation involves a sequence of push-to-the-top operations that transforms the cell ranking to represent a desired updated data. Note that the number of performed push-to-the-top operations determines when an expensive block erasure is required. However, the number of rewriting operations itself does not affect the triggering of the block erasure. Therefore, rewriting operations that require fewer push-to-the-top operations can be seen as \emph{cheaper}, and are therefore more desirable. Nevertheless, limiting the memory to cheap rewriting operations would reduce the number of potential rankings to write, and therefore would reduce the amount of information that could be stored. We refer to the number of push-to-the-top operations in a given rewriting operation as the \emph{cost of rewriting}. The study in~\cite[Section IV]{JiaMatSchBru09} considers rewriting codes with a constrained rewriting cost.
%It is shown that for a fixed cost constraint and a growing 
%We note in passing that from a Shannon-theoretical prespective, this setting can be seen as a noiseless channel-coding setting, with state information at the encoder (where the channel and state have memory and the cost constraint can be hidden in the channel model). 

%The starting point of this work is the rank-modulation rewrite-coding framework that was proposed in~\cite[Section IV]{JiaMatSchBru09}. 

The first contribution of this paper is a modification of the framework of rank-modulation rewriting codes, in two ways. First, we modify the rank-modulation scheme to allow rankings with \emph{repetitions}, meaning that multiple cells can share the same rank, where the number of cells in each rank is predetermined. And second, we extend the rewriting operation, to allow pushing a cell's level above that of any desired cell, instead of only above the level of the top cell. We justify both modifications and devise and appropriate notion of rewriting cost. 
Specifically, we define the cost to be the difference between the charge level of the \emph{highest} cell, \emph{after} the writing operation, to the charge level of the highest cell \emph{before} the rewriting operation.  
%Specifically, we define the cost to be the number of cells \emph{of different ranks} whose level \emph{after} the writing operation is higher than the level of the highest cell \emph{before} the operation. 
We suggest and explain why the new cost function compares fairly to that of the push-to-the-top model. We then go on to study rewriting codes in the modified framework.

We measure the storage rate of rewriting codes by the ratio between the number of stored information bits in each write, to the number of cells in the ranking. We study the case in which the number of cells is large (and asymptotically growing), while the cost constraint is a \emph{constant}, as this case appears to be fairly relevant for practical applications. In the model of push-to-the-top rewriting which was studied in~\cite[Section IV]{JiaMatSchBru09}, the storage rate vanishes when the number of cells grows. Our first interesting result is that the asymptotic storage rate in our modified framework converges into a \emph{positive} value (that depends on the cost constraint). Specifically, using rankings \emph{without} repetitions, i.e. the original rank modulation scheme with the modified rewriting operation, and the minimal cost constraint of a single unit, the best storage rate converges to a value of 1 bit per cell. Moreover, when ranking with repetitions is allowed, the best storage rate with a minimal cost constraint converges to a value of 2 bits per cell.

Motivated by these positive results, the rest of the paper is dedicated to the explicit construction of rank-modulation rewriting codes, together with computationally efficient encoding and decoding algorithms. The main ingredients in the code construction are recently-devised constructions of ``write-once memory'' (WOM) codes. We focus on ranking with repetitions, where both the number of cells in each rank and the number of ranks are growing. In this case, we show how to make use of capacity-achieving WOM codes to construct rank-modulation rewriting codes with an asymptotically optimal rate for any given cost constraint.

The current paper does not consider the issue of error correction. However, error-correcting codes for the rank-modulation scheme were studied extensively in recent years, as in~\cite{BarMaz10,FarSkaMil13,JiaSchBru10,TamSch10}. In addition, several variations of rank modulation were proposed and studied in \cite{EngLanSchBru11b,WanBru10}. 

The rest of the paper is organized as follows: In Section~\ref{sec:mods} we define the rank-modultion scheme and explain the proposed modifications to the scheme. In Section~\ref{sec:limits} we define the rank-modulation rewriting codes and study their information limits. Section~\ref{sec:high_level} describes the higher level of the construction we propose in this paper, and Sections \ref{sec:polar} and \ref{sec:hash} describe two alternative implementations of the building blocks of the construction. Finally, concluding remarks are provided in Section \ref{sec:conclusions}.

\section{Modifications to the Rank-Modulation Scheme}
\label{sec:mods}

In this section we motivate and define the rank-modulation scheme, together with the proposed modification to the scheme and to the rewriting process.

\subsection{Motivation for Rank Modulation}

The rank-modulation scheme is motivated by the physical and architectural properties of flash memories (and similar non-volatile memories). First, the charge injection in flash memories is a noisy process, in which an overshooting may occur. When the cells represent data by their \emph{absolute} value, such overshooting results in a different stored data than the desired one. And since the cell level cannot be decreased, the charge injection is typically performed iteratively and therefore slowly, to avoid such errors. However, in rank modulation such overshooting errors can be corrected without decreasing the cell levels, by pushing other cells to form the desired ranking. An additional issue in flash memories is the leakage of charge from the cells over time, which introduces additional errors. In rank modulation, such leakage is significantly less problematic, since it behaves similarly in spatially close cells, and thus is not likely to change the cells' ranking. A hardware implementation of the scheme was recently designed on flash memories \cite{KimParTwi12}.

We note that the motivation above is valid also for the case of ranking with repetitions, which was not considered in previous literature with respect to the rank-modulation scheme. We also note that the rank-modulation scheme in some sense reduces the amount of information that can be stored, since it limits the possible state that the cells can take. For example, it is not allowed for all the cell levels to be the same. However, this disadvantage might be worth taking for the benefits of rank modulation, and this is the case in which we are interested in this paper.

\subsection{Representing Data by Rankings with Repetitions}

In this subsection we extend the rank-modulation scheme to allow rankings with repetitions, and formally define the extended demodulation process. We refer to rankings with repetitions as \emph{permutations of multisets}, where rankings without repetitions are permutations of sets. Let $M=\{a_1^{z_1},\dots,a_q^{z_q}\}$ be a multiset of $q$ distinct elements, where each element $a_i$ appears $z_i$ times. The positive integer $z_i$ is called the multiplicity of the element $a_i$, and the cardinality of the multiset is $n=\sum_{i=1}^{q}z_i$. For a positive integer $n$, the set $\mathset{1,2,\dots,n}$ is labeled by $[n]$. We think of a permutation $\sigma$ of the multiset $M$ as a partition of the set $[n]$ into $q$ disjoint subsets, $\sigma=(\sigma(1),\sigma(2),\dots,\sigma(q))$, such that $|\sigma(i)|=z_i$ for each $i\in[q]$, and $\cup_{i\in[q]}\sigma(i)=[n]$. We also define the inverse permutation $\sigma^{-1}$ such that for each $i\in [q]$ and $j\in[n]$, $\sigma^{-1}(j)=i$ if $j$ is a member of the subset $\sigma(i)$. We label $\sigma^{-1}$ as the length-$n$ vector $\sigma^{-1}=(\sigma^{-1}(1),\sigma^{-1}(2),\dots,\sigma^{-1}(n))$. For example, if $M=\mathset{1,1,2,2}$ and $\sigma=(\mathset{1,3},\mathset{2,4})$, then $\sigma^{-1}=(1,2,1,2)$. We refer to both $\sigma$ and $\sigma^{-1}$ as a permutation, since they represent the same object.

Let $\mathfrak{S}_M$ be the set of all permutations of the multiset $M$. By abuse of notation, we view $\mathfrak{S}_M$ also as the set of the inverse permutations of the multiset $M$. For a given cardinality $n$ and number of elements $q$, it is easy to show that the number of multiset permutations is maximized if the multiplicities of all of the elements are equal. Therefore, to simplify the presentation, we take most of the multisets in this paper to be of the form $M=\{1^z,2^z,\dots,q^z\}$, and label the set $\mathfrak{S}_M$ by $\mathfrak{S}_{q,z}$. 

Consider a set of $n$ memory cells, and denote $\bfx=(x_1,x_2,\dots,x_n)\in \mathbb{R}^{n}$ as the \emph{cell-state vector}. The values of the cells represent voltage levels, but we do not pay attention to the units of these values (i.e. Volt). We represent information on the cells according to the mutiset permutation that their values induce. This permutation is derived by a demodulation process.

{\bf Demodulation:} Given positive integers $q$ and $z$, a cell-state vector $\bfx$ of length $n=qz$ is demodulated into a permutation $\pi^{-1}_{\bfx}=(\pi^{-1}_{\bfx}(1),\pi^{-1}_{\bfx}(2),\dots,\pi^{-1}_{\bfx}(n))$. Note that while $\pi^{-1}_{\bfx}$ is a function of $q,z$ and $\bfx$, $q$ and $z$ are not specified in the notation since they will be clear from the context. The demodulation is performed as follows: First, let $k_1,\ldots,k_n$ be an order of the cells such that $x_{k_1}\leq x_{k_2}\leq \cdots \leq x_{k_n}$. Then, for each $j\in[n]$, assign $\pi^{-1}_{\bfx}(k_j)=\lceil j/z\rceil$.

\begin{example}
Let $q=3$, $z=2$ and so $n=qz=6$. Assume that we wish to demodulate the cell-state vector $\bfx=(1,1.5,0.3,0.5,2,0.3)$. We first order the cells according to their values: $(k_1,k_2,\dots,k_6)=(3,6,4,1,2,5)$, since the third and sixth cells have the smallest value, and so on. Then we assign 
$$\pi^{-1}_{\bfx}(k_1=3)=\lceil 1/2\rceil=1,$$
$$\pi^{-1}_{\bfx}(k_2=6)=\lceil 2/2\rceil=1,$$
$$\pi^{-1}_{\bfx}(k_3=4)=\lceil 3/2\rceil=2,$$
 and so on, and get the permutation $\pi^{-1}_{\bfx}=(2,3,1,2,3,1)$. Note that $\pi^{-1}_{\bfx}$ is in $\mathfrak{S}_{3,2}$.
\end{example}

Note that $\pi_{\bfx}$ is not unique if for some $i\in[q]$, $x_{k_{zi}}= x_{k_{zi+1}}$. In this case, we define $\pi_{\bfx}$ to be illegal and denote $\pi_{\bfx} = F$. We label $Q_M$ as the set of all cell-state vectors that demodulate into a valid permutation of $M$. That is, $Q_M=\{\bfx \in \mathbb{R}^n \ | \pi_{\bfx} \neq F \}$. So for all $\bfx\in Q_M$ and $i\in [q]$, we have $x_{k_{zi}}< x_{k_{zi+1}}$.
For $j\in[n]$, the value $\pi^{-1}(j)$ is called the \emph{rank} of cell $j$ in the permutation $\pi$. 

\subsection{Rewriting in Rank Modulation}

In this subsection we extend the rewriting operation in the rank-modulation scheme. Previous work considered a writing operation called ``push-to-the-top'', in which a certain cell is pushed to be the highest in the ranking~\cite{JiaMatSchBru09}. Here we suggest to allow to push a cell to be higher than the level of \emph{any} specific other cell. We note that this operation is still resilient to overshooting errors, and therefore benefits from the advantage of fast writing, as the push-to-the-top operations.

We model the flash memory such that when a user wishes to store a message on the memory, the cell levels can only increase. When the cells reach their maximal levels, an expensive erasure operation is required. Therefore, in order to maximize the number of writes between erasures, it is desirable to raise the cell levels as little as possible on each write. 
For a cell-state vector $\bfx\in Q_M$, denote by $\Gamma_{\bfx}(i)$ the highest level among the cells with rank $i$ in $\pi_{\bfx}$. That is, 
\[\Gamma_{\bfx}(i)=\max_{j\in \pi_{\bfx}(i)}\{x_j\}.\]
Let $\bfs$ be the cell-state vector of the memory before the writing process takes place, and let $\bfx$ be the cell-state vector after the write. 
In order to reduce the possibility of error in the demodulation process, a certain gap must be placed between the levels of cells with different ranks. Since the cell levels's units are somewhat arbitrary, we set this gap to be the value $1$, for convenience.
The following modulation method minimizes the increase in the cell levels.

{\bf Modulation:} 
Writing a permutation $\pi$ on a memory with state $\bfs$. The output is the new memory state, denoted by $\bfx$.
\begin{enumerate}
\item For each $j\in \pi(1)$, assign $x_j\Leftarrow s_j$.
\item For $i=2,3,\dots,q$, for each $j\in \pi(i)$, assign
$$x_j\Leftarrow\max\{s_j,\Gamma_{\bfx}(i-1)+1\}.$$
\end{enumerate}

\begin{example}
Let $q=3$, $z=2$ and so $n=qz=6$. Let the state be $\bfs=(2.7,4,1.5,2.5,3.8,0.5)$ and the target permutation be $\pi^{-1}=(1,1,2,2,3,3)$. In step 1 of the modulation process, we notice that $\pi(1)=\mathset{1,2}$ and so we set 
$$x_1\Leftarrow s_1=2.7$$
and 
$$x_2\Leftarrow s_2=4.$$
In step 2 we have $\pi(2)=\mathset{3,4}$ and $\Gamma_{\bfx}(1)=\max\mathset{x_1,x_2}=\max\mathset{2.7,4}=4$, so we set 
$$x_3\Leftarrow\max\mathset{s_3,\Gamma_{\bfx}(1)+1}=\max\mathset{1.5,5}=5$$
and 
$$x_4\Leftarrow\max\mathset{s_4,\Gamma_{\bfx}(1)+1}=\max\mathset{2.5,5}=5.$$
And in the last step we have $\pi(3)=\mathset{5,6}$ and $\Gamma_{\bfx}(2)=5$, so we set 
$$x_5\Leftarrow\max\mathset{3.8,6}=6$$
and 
$$x_6\Leftarrow\max\mathset{0.5,6}=6.$$
In summary, we get $\bfx=(2.7,4,5,5,6,6)$, which demodulates into $\pi_{\bfx}^{-1}=(1,1,2,2,3,3)=\pi^{-1}$, as required.
\end{example}

Since the cell levels cannot decrease, we must have $x_j\ge s_j$ for each $j\in[n]$. In addition, for each $j_1$ and $j_2$ in $[n]$ for which $\pi^{-1}(j_1)>\pi^{-1}(j_2)$, we must have $x_{j_1}>x_{j_2}$. Therefore, the proposed modulation process minimizes the increase in the levels of all the cells.

\section{Definition and Limits of Rank-Modulation Rewriting Codes}
\label{sec:limits}

Remember that the level $x_j$ of each cell is upper bounded by a certain value. Therefore, given a state $\bfs$, certain permutations $\pi$ might require a block erasure before writing, while others might not. In addition, some permutations might get the memory state \emph{closer} to a state in which an erasure is required than other permutations. In order to maximize the number of writes between block erasures, we add redundancy by letting \emph{multiple} permutations represent the \emph{same} information message. This way, when a user wishes to store a certain message, she could choose one of the permutations that represent the required message such that the chosen permutation will increase the cell levels in the least amount. Such a method can increase the longevity of the memory in the expense of the amount of information stored on each write. The mapping between the permutations and the messages they represent is called a rewriting code.

To analyze and design rewriting codes, we focus on the difference between $\Gamma_{\bfx}(q)$ and $\Gamma_{\bfs}(q)$. Using the modulation process we defined above, the vector $\bfx$ is a function of $\bfs$ and $\pi$, and therefore the difference $\Gamma_{\bfx}(q)-\Gamma_{\bfs}(q)$ is also a function of $\bfs$ and $\pi$. We label this difference by $\alpha(\bfs\to\pi)=\Gamma_{\bfx}(q)-\Gamma_{\bfs}(q)$ and call it the \emph{rewriting cost}, or simply the cost. We motivate this choice by the following example. Assume that the difference between the maximum level of the cells and $\Gamma_{\bfs}(q)$ is 10 levels. Then only the permutations $\pi$ which satisfy $\alpha(\bfs\to\pi)\le 10$ can be written to the memory without erasure. Alternatively, if we use a rewriting code that guarantees that for any state $\bfs$, any message can be stored with, say, cost no greater than 1, then we can guarantee to write 10 more times to the memory before an erasure will be required. Such rewriting codes are the focus of this paper.

The cost $\alpha(\bfs\to\pi)$ is defined according to the vectors $\bfs$ and $\bfx$. However, it will be helpful for the study of rewriting codes to have some understanding of the cost in terms of the demodulation of the state $\bfs$ and the permutation $\pi$. To establish such connection, we assume that the state $\bfs$ is a result of a previous modulation process. This assumption is reasonable, since we are interested in the scenario of multiple successive rewriting operations. In this case, for each $i\in[q-1]$, $\Gamma_{\bfs}(i+1)-\Gamma_{\bfs}(i)\ge1$, by the modulation process.
Let $\sigma_{\bfs}$ be the permutation obtained from the demodulation of the state $\bfs$. We present the connection in the following proposition.

\begin{proposition}
\label{prop:cost}
Let $M$ be a multiset of cardinality $n$. If $\Gamma_{\bfs}(i+1)-\Gamma_{\bfs}(i)\ge1$ for all $i\in[q-1]$, and $\pi$ is in $\mathfrak{S}_M$, then
\begin{equation}
\label{eq:cost}
\alpha(\bfs\to\pi)\le\max_{j\in [n]}\{\sigma_{\bfs}^{-1}(j)-\pi^{-1}(j)\}
\end{equation}
with equality if $\Gamma_q(\bfs)-\Gamma_1(\bfs)=q-1$.
\end{proposition}

The proof of Proposition \ref{prop:cost} is brought in Appendix \ref{app:cost}. We would take a worst-case approach, and opt to design codes that guarantee that on each rewriting, the value $\max_{j\in [n]}\{\sigma_{\bfs}^{-1}(j)-\pi^{-1}(j)\}$ is bounded. For permutations $\sigma$ and $\pi$ in $\mathfrak{S}_{q,z}$, the rewriting cost $\alpha(\sigma\to\pi)$ is defined as 
\begin{equation}
\label{eq:cost_perm}
\alpha(\sigma\to\pi)=\max_{j\in [n]}\{\sigma^{-1}(j)-\pi^{-1}(j)\}.
\end{equation}
This expression is an asymmetric version of the Chebyshev distance (also known as the $L_{\infty}$ distance). For simplicity, we assume that the channel is noiseless and don't consider the error-correction capability of the codes. However, such consideration would be essential for practical applications. %Since the channel is noiseless, the code is uniquely defined by the decoding map. We are now ready to define \emph{rank-modulation rewriting codes}, the main objects of study of this paper. 

\subsection{Definition of Rank-Modulation Rewriting Codes}

A rank-modulation rewriting code is a partition of the set of multiset permutations, such that each part represents a different information message, and each message can be written on each state with a cost that is bounded by some parameter $r$. A formal definition follows.

\begin{definition}
\label{def:rmrc}
{\bf (Rank-modulation rewriting codes)} Let $q,z,r$ and $K_R$ be positive integers, and let $\cC$ be a subset of $\mathfrak{S}_{q,z}$ called the codebook. Then a surjective function $D_R:\cC\to[K_R]$ is a $(q,z,K_R,r)$ \textbf{rank-modulation rewriting code (RM rewriting code)} if for each message $m\in[K_R]$ and state $\sigma\in \cC$, there exists a permutation $\pi$ in $D_R^{-1}(m)\subseteq\cC$ such that $\alpha(\sigma\to\pi)\le r$.
\end{definition}

$D_R^{-1}(m)$ is the set of permutations that represent the message $m$. It could also be insightful to study rewriting codes according to an average cost constraint, assuming some distribution on the source and/or the state. However, we use the wort-case constraint since it is easier to analyze. The amount of information stored with a $(q,z,K_R,r)$ RM rewriting code is $\log K_R$ bits (all of the logarithms in this paper are binary). Since it is possible to store up to $\log |\mathfrak{S}_{q,z}|$ bits with permutations of a multiset $\mathset{1^z,\dots,q^z}$, it could be natural to define the code rate as:
$$R'=\frac{\log K_R}{\log|\mathfrak{S}_{q,z}|}.$$
However, this definition doesn't give much engineering insight into the amount of information stored in a set of memory cells. Therefore, we define the rate of the code as the amount of information stored per  memory cell:
$$R=\frac{\log K_R}{qz}.$$
An encoding function $E_R$ for a code $D_R$ maps each pair of message $m$ and state $\sigma$ into a permutation $\pi$ such that $D_R(\pi)=m$ and $\alpha(\sigma\to\pi)\le r$. By abuse of notation, let the symbols $E_R$ and $D_R$ represent both the functions and the algorithms that compute those functions. If $D_R$ is a RM rewriting code and $E_R$ is its associated encoding function, we call the pair $(E_R,D_R)$ a rank-modulation rewrite coding scheme.

Rank-modulation rewriting codes were proposed by Jiang \emph{et al.} in \cite{JiaMatSchBru09}, in a more restrictive model than the one we defined above. The model in \cite{JiaMatSchBru09} is more restrictive in two senses. First, the mentioned model used the rank-modulation scheme with permutations of sets only, while here we also consider permutations of multisets. And second, the rewriting operation in the mentioned model was composed only of a cell programming operation called ``push to the top", while here we allow a more opportunistic programming approach. A push-to-the-top operation raises the charge level of a single cell above the rest of the cells in the set. As described above, the model of this paper allows to raise a cell level above a subset of the rest of the cells. The rate of RM rewriting codes with push-to-the-top operations and cost of $r=1$ tends to zero with the increase in the block length $n$. On the contrary, we will show that the rate of RM rewriting codes with cost $r=1$ and the model of this paper tends to 1 bit per cell with permutations of sets, and 2 bits per cell with permutations of multisets.

\subsection{Limits of Rank-Modulation Rewriting Codes}

For the purpose of studying the limits of RM rewriting codes, we define the ball of radius $r$ around a permutation $\sigma$ in $\mathfrak{S}_{q,z}$ by
$$B_{q,z,r}(\sigma)=\{\pi\in \mathfrak{S}_{q,z}|\alpha(\sigma\to\pi)\le r\},$$
and derive its size in the following lemma.

\begin{lemma}
\label{lem:ball_size}
For positive integers $q$ and $z$, if $\sigma$ is in $\mathfrak{S}_{q,z}$ then
$$|B_{q,z,r}(\sigma)|=\binom{(r+1)z}{z}^{q-r}\prod_{i=1}^{r}\binom{iz}{z}.$$
\end{lemma}

\begin{IEEEproof}
Let $\pi\in B_{q,z,r}(\sigma)$. By the definition of $B_{q,z,r}(\sigma)$, for any $j\in \pi(1)$, $\sigma^{-1}(j)-1\le r$, and thus $\sigma^{-1}(j)\le r+1$. Therefore, there are $\binom{(r+1)z}{z}$ possibilities for the set $\pi(1)$ of cardinality $z$. Similarly, for any $i\in \pi(2)$, $\sigma(i)^{-1}\le r+2$. So for each fixed set $\pi(1)$, there are $\binom{(r+1)z}{z}$ possibilities for $\pi(2)$, and in total $\binom{(r+1)z}{z}^2$ possibilities for the pair of sets $(\pi(1),\pi(2))$. The same argument follows for all $i\in[q-r]$, so there are $\binom{(r+1)z}{z}^{q-r}$ possibilities for the sets $(\pi(1),\dots,\pi(q-r))$. The rest of the sets of $\pi$: $\pi(q-r+1),\pi(q-r+2),\dots,\pi(q)$, can take any permutation of the multiset $\mathset{(q-r+1)^z,(q-r+2)^z,\dots,q^z}$, giving the statement of the lemma.
\end{IEEEproof}

Note that the size of $B_{q,z,r}(\sigma)$ is actually \emph{not} a function of $\sigma$. Therefore we denote it by $|B_{q,z,r}|$. 

\begin{proposition}
\label{lem:bound}
Let $D_R$ be a $(q,z,K_R,r)$ RM rewriting code. Then 
$$K_R\le|B_{q,z,r}|.$$
\end{proposition}

\begin{IEEEproof}
Fix a state $\sigma\in \cC$. By Definition \ref{def:rmrc} of RM rewriting codes, for any message $m\in[K_R]$ there exists a permutation $\pi$ such that $D_R(\pi)=m$ and $\pi$ is in $B_{q,z,r}(\sigma)$. It follows that $B_{q,z,r}(\sigma)$ must contain $K_R$ different permutations, and so its size must be at least $K_R$.
\end{IEEEproof}

\begin{corollary}
\label{cor:limit}
Let $R(r)$ be the rate of an $(q,z,K_R,r)$-RM rewriting code. Then 
\[R(r)<(r+1)H\left(\frac{1}{r+1}\right),\]
where $H(p)=-p\log p-(1-p)\log(1-p)$ . In particular, $R(1)<2$.
\end{corollary}

\begin{IEEEproof}
\begin{align*}
\log |B_{q,z,r}|=&\sum_{i=1}^{r}\log\binom{iz}{z}+(q-r)\log\binom{(r+1)z}{z}\\
<&r\log\binom{(r+1)z}{z}+(q-r)\log\binom{(r+1)z}{z}\\
= &q\log\binom{(r+1)z}{z}\\
< &q\cdot (r+1)z H\left(\frac{1}{r+1}\right),
\end{align*}
where the last inequality follows from Stirling's formula.
So we have
$$R(r)=\frac{\log K_R}{qz}\le \frac{\log |B_{q,z,r}|}{qz}<(r+1)H\left(\frac{1}{r+1}\right).$$
The case of $r=1$ follows immediately.
\end{IEEEproof}

We will later show that this bound is in fact tight, and therefore we call it the \emph{capacity} of RM rewriting codes and denote it as 
$$C_R(r)=(r+1)H\left(\frac{1}{r+1}\right).$$
Henceforth we omit the radius $r$ from the capacity notation and denote it by $C_R$. To further motivate the use of multiset permutations rather than set permutation, we can observe the following corollary.

\begin{corollary}
\label{cor:limit_set}
Let $R(r)$ be the rate of an $(q,1,K_R,r)$-RM rewriting code. Then $R(r)<\log(r+1)$, and in particular, $R(1)<1$.
\end{corollary}

\begin{IEEEproof}
Note first that $|B_{q,z,r}|=(r+1)^{q-r}r!$. So we have
\begin{align*}
\log |B_{q,z,r}|=&\log r!+(q-r)\log(r+1)\\
<&r\log (r+1)+(q-r)\log(r+1)\\
= &q\log(r+1).
\end{align*}
Therefore,
$$R(r)\le\frac{\log |B_{q,z,r}|}{q}<\log(r+1),$$
and the case of $r=1$ follows immediately.
\end{IEEEproof}

In the case of $r=1$, codes with multiset permutations could approach a rate close to 2 bits per cell, while there are no codes with set permutations and rate greater than 1 bit per cell. The constructions we present in this paper are analyzed only for the case of multiset permutations with a large value of $z$. We now define two properties that we would like to have in a family of RM rewrite coding schemes. First, we would like the rate of the codes to approach the upper bound of Corollary \ref{cor:limit}. We call this property \emph{capacity achieving}.

\begin{definition}
\label{def:capacity_achieving}
{\bf (Capacity-achieving family of RM rewriting codes)} For a positive integer $i$, let the positive integers $q_i,z_i$ and $K_i$ be some functions of $i$, and let $n_i=q_i z_i$ and $R_i=(1/n_i)\log K_i$.
Then an infinite family of $(q_i,z_i,K_i,r)$ RM rewriting codes is called \textbf{capacity achieving} if 
$$\lim_{i\to\infty}R_i=C_R.$$
\end{definition}

The second desired property is computational efficiency. We say that a family of RM rewrite coding schemes $(E_{R,i},D_{R,i})$ is \emph{efficient} if the algorithms $E_{R,i}$ and $D_{R,i}$ run in polynomial time in $n_i=q_iz_i$. The main result of this paper is a construction of an efficient capacity-achieving family of RM rewrite coding schemes.

\section{High-Level Construction}
\label{sec:high_level}

The proposed construction is composed of two layers. The higher layer of the construction is described in this section, and two alternative implementations of the lower layer are described in the following two sections. The high-level construction involves several concepts, which we introduce one by one. The first concept is to divide the message into $q-r$ parts, and encode and decode each part separately. The codes that are used for the different message parts are called ''ingredient codes''. We demonstrate this concept in Subsection~\ref{ss:q3z2r1} by an example in which $q=3$,$z=2$ and $r=1$, and the RM code is divided into $q-r=2$ ingredient codes.

The second concept involves the implementation of the ingredient codes when the parameter $z$ is greater than 2. We show that in this case the construction problem reduces to the construction of the so-called ``constant-weight WOM codes''. We demonstrate this in Subsection~\ref{ss:gen_z} with a construction for general values of $z$, where we show that capacity-achieving constant-weight WOM codes lead to capacity achieving RM rewriting codes. Next, in Subsections~\ref{ss:gen_q} and~\ref{ss:gen_r}, we generalize the parameters $q$ and $r$, where these generalizations are conceptually simpler. 

Once the construction is general for $q$, $z$ and $r$, we modify it slightly in Subsection~\ref{ss:weak} to accommodate a weaker notion of WOM codes, which are easier to construct. The next two sections present two implementations of capacity-achieving weak WOM codes, that can be used to construct capacity-achieving RM rewriting codes.

A few additional definitions are needed for the description of the construction. First, let $2^{[n]}$ denote the set of all subsets of $[n]$. Next, let the function $\theta_n: 2^{[n]} \to \{0,1\}^n$ be defined such that for a subset $S\subseteq [n]$, $\theta_n(S)=(\theta_{n,1},\theta_{n,2},\dots,\theta_{n,n})$ is its characteristic vector, where
\[ \theta_{n,j} = \left\{ 
  \begin{array}{l l}
    0 & \quad \text{if $j\notin S$}\\
    1 & \quad \text{if $j\in S$.}
  \end{array} \right.\]
For a vector $\bfx$ of length $n$ and a subset $S\subseteq [n]$, we denote by $\bfx_S$ the vector of length $|S|$ which is obtained by "throwing away" all the positions of $\bfx$ outside of $S$. For positive integers $n_1\le n_2$, the set $\mathset{n_1,n_1+1,\dots,n_2}$ is labeled by $[n_1:n_2]$. Finally, for a permutation $\sigma\in\mathfrak{S}_{q,z}$, we define the set $U_{i_1,i_2}(\sigma)$ as the union of the sets $\mathset{\sigma(i)}_{i\in[i_1:i_2]}$ if $i_1\le i_2$. If $i_1>i_2$, we define $U_{i_1,i_2}(\sigma)$ to be the empty set. 

%\begin{example}
%We demonstrate a construction of a $(q=3,z=2,K_R=30,r=1)$ RM rewrite coding scheme. The number of cells is $n=qz=6$, and the number permutations of the multiset $\{1,1,2,2,3,3\}$ is $6!/2^3=90$. 
%
%\end{example}

%To facilitate the explanation of the construction, we deal first with the case in which the parameter $r$, the bound on the cost, is equal to 1. The ideas will then generalize naturally to any value of $r$. 
%To construct a $(q,z,K_R,r)$ RM rewriting scheme $\mathset{E_R,D_R}$, we will think first of the encoder $E_R$. Given a message $m\in[K_R]$ and a state $\sigma$ in the codebook $\cC$, the encoder needs to find a permutation $\pi$ in the ball $B_{q,z,r}(\sigma)$, such that  $D_R(\pi)=m$, meaning that the permutation $\pi$ represents the message $m$. For $\pi$ to be in the ball $B_{q,z,1}(\sigma)$, the cost $\alpha(\sigma\to\pi)$ must not be greater that $r$. By the definition of the cost, this implies that for each cell $j\in[n]$, we must have $\sigma^{-1}(j)-\pi^{-1}(j)\le r$. 

\subsection{A Construction for $q=3$,$z=2$ and $r=1$}
\label{ss:q3z2r1}

%To gain an intuition about the code construction, we start by considering an example in which the number of ranks is $q=3$, the number of cells in each rank is $z=2$, and the cost constraint is $r=1$. 

In this construction we introduce the concept of dividing the code into multiple ingredient codes. The motivation for this concept comes from a view of the encoding process as a sequence of choices. Given a message $m$ and a state permutation $\sigma$, the encoding process needs to find a permutation $\pi$ that represents $m$, such that the cost $\alpha(\sigma\to\pi)$ is no greater then the cost constraint $r$. The cost function $\alpha(\sigma\to\pi)$ is defined in Equation~\ref{eq:cost_perm} as the maximal \emph{drop} in rank among the cells, when moving from $\sigma$ to $\pi$. In other words, we look for the cell that dropped the most amount of ranks from $\sigma$ to $\pi$, and the cost is the number of ranks that this cell has dropped. If cell $j$ is at rank 3 in $\sigma$ and its rank is changed to 1 in $\pi$, it dropped 2 ranks. In our example, since $q=3$, a drop of 2 ranks is the biggest possible drop, and therefore, if at least one cell dropped by 2 ranks, the rewriting cost would be 2.

In the setting of $q=3$ ranks, $z=2$ cells per rank, and cost constraint of $r=1$, to make sure that a the rewriting cost would not exceed 1, it is enough to ensure that the 2 cells of rank 3 in $\sigma$ do not drop into rank 1 in $\pi$. So the cells that take rank 1 in $\pi$ must come from ranks 1 or 2 in $\sigma$. This motivates us to look at the encoding process as a sequence of 2 decisions. First, the encoder chooses two cells out of the 4 cells in ranks 1 and 2 in $\sigma$, to occupy rank 1 in $\pi$. 
%The set $\pi(1)$ (the set of cells with rank 1 in $\pi$) must be chosen such that there will be at least one permutation $\pi$ in the code with the chosen $\pi(1)$, that represents the message $m$. 
Next, after the $\pi(1)$ (the set of cells with rank 1 in $\pi$) is selected, the encoder completes the encoding process by choosing a way to arrange the remaining 4 cells in ranks 2 and 3 of $\pi$. There are $\binom{4}{2}=6$ such arrangements, and they all satisfy the cost constraint, since a drop from a rank no greater than 3 into a rank no smaller than 2 cannot exceed a magnitude of 1 rank. So the encoding process is split into two decisions, which define it entirely. 

%Following this process, the rewriting cost is guaranteed to be at most 1. The encoder must make these 2 choices such that the resulting permutation $\pi$ represents the message $m$.

The main concept in this subsection is to think of the message as a pair $\bfm=(m_1,m_2)$, such that the first step of the encoding process encodes $m_1$, and the second step encodes $m_2$. The first message part, $m_1$, is encoded by the set $\pi(1)$. 
%Each selection of 2 cells for $\pi(1)$ is called a codeword.
 To satisfy the cost constraint of $r=1$, the set $\pi(1)$ must be chosen from the 4 cells in ranks 1 and 2 in $\sigma$. These 4 cells are denoted by $U_{1,2}(\sigma)$.  
For each $m_1$ and set $U_{1,2}(\sigma)$, the encoder needs to find 2 cells from $U_{1,2}(\sigma)$ that represent $m_1$. Therefore, there must be \emph{multiple} selections of 2 cells that represent $m_1$.

The encoding function for $m_1$ is denoted by $E_W(m_1,U_{1,2}(\sigma))$, and the corresponding decoding function is denoted by $D_W(\pi(1))$. We denote by $D_W^{-1}(m_1)$ the \emph{set} of subsets that $D_W$ maps into $m_1$. We denote the number of possible values that $m_1$ can take by $K_W$. To demonstrate the code $D_W$ for $m_1$, we show an example that contains $K_W=5$ messages.

\begin{example}
\label{ex:small_wom}
Consider the following code $D_W$, defined by the values of $D_W^{-1}$:
\begin{align*}
D_W^{-1}(1)=&\Big\{\{1,2\},\{3,4\},\{5,6\}\Big\}\\
D_W^{-1}(2)=&\Big\{\{1,3\},\{2,6\},\{4,5\}\Big\}\\
D_W^{-1}(3)=&\Big\{\{1,4\},\{2,5\},\{3,6\}\Big\}\\
D_W^{-1}(4)=&\Big\{\{1,5\},\{2,3\},\{4,6\}\Big\}\\
D_W^{-1}(5)=&\Big\{\{1,6\},\{2,4\},\{3,5\}\Big\}.
\end{align*}
To understand the code, assume that $m_1=3$ and $\sigma^{-1}=(1,2,1,3,2,3)$, so that the cells of ranks 1 and 2 in $\sigma$ are $U_{1,2}(\sigma)=\{1,2,3,5\}$. The encoder needs to find a set in $D_W^{-1}(3)$, that is a subset of $U_{1,2}(\sigma)=\{1,2,3,5\}$. In this case, the only such set is $\{2,5\}$. So the encoder chooses cells 2 and 5 to occupy rank 1 of $\pi$, meaning that the rank of cells 2 and 5 in $\pi$ is 1, or that $\pi(1)=\{2,5\}$. To find the value of $m_1$, the decoder calculates the function $D_W(\pi(1))=3$. It is not hard to see that for any values of $m_1$ and $U_{1,2}(\sigma)$ (that contains 4 cells), the encoder can find 2 cells from $U_{1,2}(\sigma)$ that represent $m_1$.
\end{example}

 The code for $m_2$ is simpler to design. The encoder and decoder both know the identity of the 4 cells in ranks 2 and 3 of $\pi$, so each arrangement of these two ranks can correspond to a different message part $m_2$. We denote the number of messages in the code for $m_2$ by $K_M$, and define the multiset $M=\{2,2,3,3\}$. We also denote the pair of sets $(\pi(2),\pi(3))$ by $\pi_{[2:3]}$. Each arrangement of $\pi_{[2:3]}$ corresponds to a different permutation of $M$, and encodes a different message part $m_2$. So we let 
$$K_M=|\mathfrak{S}_M|=\binom{4}{2}=6.$$

For simplicity, we encode $m_2$ according to the \emph{lexicographic} order of the permutations of $M$. For example, $m_2=1$ is encoded by the permutation $(2,2,3,3)$, $m_2=2$ is encoded by $(2,3,2,3)$, and so on. If, for example, the cells in ranks 2 and 3 of $\pi$ are $\{1,3,4,6\}$, and the message part is $m_2=2$, the encoder sets 
$$\pi_{[2:3]}=(\pi(2),\pi(3))=(\{1,4\},\{3,6\}).$$ The bijective mapping form $[K_M]$ to the permutations of $M$ is denoted by $h_M(m_2)$, and the inverse mapping by $h^{-1}(\pi_{[2:3]})$. The code $h_M$ is called an enumerative code.

The message parts $m_1$ and $m_2$ are encoded sequentially, but can be decoded in parallel. 
The number of messages that the RM rewriting code in this example can store is 
$$K_R=K_W\times K_M=5\times 6=30,$$
as each rank stores information independently.  

\begin{construction}
\label{con:q3z2r1}
Let $K_W=5,q=3,z=2,r=1$, let $n=qz=6$ and let $(E_W,D_{W})$ be defined according to Example~\ref{ex:small_wom}. Define the multiset $M=\mathset{2,2,3,3}$ and let $K_M=|\mathfrak{S}_M|=6$ and $K_R=K_W\cdot K_M=30$. The codebook $\cC$ is defined to be the entire set $\mathfrak{S}_{3,2}$. A $(q=3,z=2,K_R=30,r=1)$ RM rewrite coding scheme $\mathset{E_R,D_R}$ is constructed as follows:

The \textbf{encoding} algorithm $E_R$ receives a message $\bfm=(m_1,m_2)\in[K_W]\times[K_M]$ and a state permutation $\sigma\in \mathfrak{S}_{3,2}$, and returns a permutation $\pi$ in $B_{3,2,1}(\sigma)\cap D^{-1}_R(\bfm)$ to store in the memory. It is constructed as follows:
\begin{algorithmic}[1]
\STATE $\pi(1)\Leftarrow E_{W}(m_1,U_{1,2}(\sigma))$
\STATE $\pi_{[2:3]}\Leftarrow h_M(m_{2})$
\end{algorithmic}

The \textbf{decoding} algorithm $D_R$ receives the stored permutation $\pi\in \mathfrak{S}_{3,2}$, and returns the stored message $\bfm=(m_1,m_2)\in[K_W]\times[K_M]$. It is constructed as follows:
\begin{algorithmic}[1]
\STATE $m_1\Leftarrow D_W(\pi(1))$
\STATE $m_{2}\Leftarrow h_M^{-1}(\pi_{[2:3]})$
\end{algorithmic}
\end{construction} 

The rate of the code $D_R$ is 
$$R_R=(1/n)\log_2(K_R)=(1/6)\log(30)\approx 0.81.$$
The rate can be increased up to 2 bits per cell while keeping $r=1$, by increasing $z$ and $q$. We continue by increasing the parameter $z$.

\subsection{Generalizing the Parameter $z$}
\label{ss:gen_z}

In this subsection we generalize the construction to arbitrary values for the number of cells in each rank, $z$. The code for the second message part, $h_M$, generalizes naturally for any value of $z$, by taking $M$ to be the multiset $M=\{2^z,3^z\}$. Since $z$ now can be large, it is important to choose the bijective functions $h_M$ and $h^{-1}_M$ such that they could be computed efficiently. Luckily, several such efficient schemes exist in the literature, such as the scheme described in~\cite{MilVas00}.

 The code $D_W$ for the part $m_1$, on the contrary, does not generalize naturally, since $D_W$ in Example~\ref{ex:small_wom} does not have a natural generalization. To obtain such a generalization, we think of the characteristic vectors of the subsets of interest. 
%The code we need for a general $z$ is a partition $D_W^{-1}$ of the subsets of size $z$ of $[n]$ into $K_W$ parts, such that for each subset $U_{1,2}(\sigma)$ of $[n]$ of size $2z$, each part of $D^{-1}_W$ contain at least one subset of $U_{1,2}(\sigma)$.
% To construct such a general scheme, we observe that this problem is similar to a concept know in the literature as Write-Once Memory codes, or WOM codes. To make this connection, we think of the subsets in $D_W^{-1}$ in term of their characteristic vectors. The codes that we need correspond to WOM codes in which each codeword has a constant Hamming weight (number of non-zero bits), since we the subsets we need have a constant size. Therefore we will call these codes \emph{Constant-Weight WOM codes}.
%To tighten the connection to WOM codes, we introduce a few additional notation.
The characteristic vector of $U_{1,2}(\sigma)$ is denoted as $\bfs=\theta_n(U_{1,2}(\sigma))$ (where  $n=qz$), and is referred to as the state vector. The vector $\bfx=\theta_n(\pi(1))$ is called the codeword. The constraint $\pi(1)\subset U_{1,2}(\sigma)$ is then translated to the constraint $\bfx\le\bfs$, which means that for each $j\in [n]$ we must have $x_j\le s_j$. We now observe that this coding problem is similar to a concept known in the literature as Write-Once Memory codes, or WOM codes (see, for example,~\cite{RivSha82,YaaKaySieVarWol12}). In fact, the codes needed here are WOM codes for which the Hamming weight (number of non-zero bits) of the codewords is constant. Therefore, we say that $D_W$ needs to be a ``constant-weight WOM code''.
%Write-once memories are typically defined such that each cell can be written once from 0 to 1. However, here it is more convenient to consider cells that can be written once from 1 to 0, since then a cell in state $s_j=0$ cannot be changed anymore, and we must have $x_j=0$. This is equivalent to the notation $x_j\le s_j$. %The fact that we "reverse" the definition of write-once memories will of course not prevent us from using any result on WOM codes from the literature. 
%Since the size of the set $\pi(1)$ is equal to $z$, the WOM codeword $\bfx$ must have a Hamming weight of $z$. 
We use the word `weight' from now on to denote the Hamming weight of a vector. %Therefore, we will be interested in WOM codes in which all of the codewords have the same weight (henceforth the term weight refers to the Hamming weight). Such codes were not considered in the literature, and we naturally call them \emph{constant-weight WOM codes}. 
%For the definition of constant-weight WOM codes we need an additional notation.

We define next the requirements of $D_W$ in a vector notation. For a positive integer $n$ and a real number $w\in[0,1]$, we let $J_{w}(n)\subset\{0,1\}^n$ be the set of all vectors of $n$ bits whose weight equals $\lfloor wn\rfloor$. We use the name ``constant-weight \emph{strong} WOM code'', since we will need to use a weaker version of this definition later. The weight of $\bfs$ in $D_W$ is $2n/3$, and the weight of $\bfx$ is $n/3$. However, we allow for more general weight in the following definition, in preperation for the generalization of the number of ranks, $q$. 

\begin{definition} 
\label{def:cwswom}
{\bf (Constant-weight strong WOM codes)} Let $K_W$ and $n$ be positive integers and let $w_s$ be a real number in $[0,1]$ and $w_x$ be a real number in $[0,w_s]$. A surjective function $D_W:J_{w_x}(n)\to[K_W]$ is an $(n,K_W,w_s,w_x)$ \textbf{constant-weight strong WOM code} if for each message $m\in[K_W]$ and state vector $\bfs\in J_{w_s}(n)$, there exists a codeword vector $\bfx\le\bfs$ in the subset $D_W^{-1}(m)\subseteq J_{w_x}(n)$. The rate of a constant-weight strong WOM code is defined as $R_W=(1/n)\log K_W$.
\end{definition}

The code $D_W$ in Example~\ref{ex:small_wom} is a $(n=6,K_W=5, w_s=2/3,w_x=1/3)$ constant-weight strong WOM code. It is useful to know the tightest upper bound on the rate of constant-weight strong WOM codes, which we call the capacity of those codes.

\begin{proposition}
\label{prop:wom_capacity}
Let $w_s$ and $w_x$ be as defined in Definition \ref{def:cwswom}. Then the capacity of constant-weight strong WOM codes is 
$$C_W=w_sH(w_x/w_s).$$
\end{proposition}

The proof of Proposition \ref{prop:wom_capacity} is brought in Appendix \ref{app:wom_capacity}. 

%\begin{example}
%Let $K_W=3$, $n=4$, $w_s=3/4$ and $w_x=1/2$. We define the following constant-weight strong WOM code:
%\begin{align*}
%D_W^{-1}(1)&=\{0011,1100\}\\
%D_W^{-1}(2)&=\{0101,1010\}\\
%D_W^{-1}(3)&=\{0110,1001\}
%\end{align*}
%For a message $m=2$ and a state $\bfs=1101\in J_{3/4}(4)$, for example, the codeword $\bfx=0101\le1101=\bfs$ is in the subset $D_W^{-1}(2)$. We can verify that for each message and state, there exists an appropriate codeword in this code.
%\end{example}

We also define the notions of coding scheme, capacity achieving and efficient family for constant-weight strong WOM codes in the same way we defined it for RM rewriting codes. To construct capacity-achieving RM rewriting codes, we will need to use capacity-acheving constant-weight WOM codes as ingredients codes. However, we do not know how to construct an efficient capacity-achieving family of constant-weight strong WOM coding schemes. Therefore, we will present later a weaker notion of WOM codes, and show how to use it for the construction of RM rewriting codes.

\subsection{Generalizing the Number of Ranks $q$}
\label{ss:gen_q}

We continue with the generalization of the construction, where the next parameter to generalize is the number of ranks $q$. So the next scheme has general parameters $q$ and $z$, while the cost constraint $r$ is still kept at $r=1$. In this case, we divide the message into $q-1$ parts, $m_1$ to $m_{q-1}$. The encoding now starts in the same way as in the previous case, with the encoding of the part $m_1$ into the set $\pi(1)$, using a constant-weight strong WOM code. However, the parameters of the WOM code need to be slightly generalized. The numbers of cells now is $n=qz$, and $E_W$ still chooses $z$ cells for rank 1 of $\pi$ out of the $2z$ cells of ranks 1 and 2 of $\sigma$. So we need a WOM code with the parameters $w_s=2/q$ and $w_x=1/q$.

The next step is to encode the message part $m_2$ into rank 2 of $\pi$. We can perform this encoding using the same WOM code $D_W$ that was used for $m_1$. However, there is a difference now in the identity of the cells that are considered for occupying the set $\pi(2)$. In $m_1$, the cells that were considered as candidates to occupy $\pi(1)$ were the $2z$ cells in the set $U_{1,2}(\sigma)$, since all of these cell could be placed in $\pi(1)$ without dropping their rank (from $\sigma$ to $\pi$) by more then 1. In the encoding of $m_2$, we choose cells for rank 2 of $\pi$, so the $z$ cells from rank 3 of $\sigma$ can also be considered. Another issue here is that the cells that were already chosen for rank 1 of $\pi$ should \emph{not} be considered as candidates for rank 2. Taking these consideration into account, we see that the candidate cells for $\pi(2)$ are the $z$ cells that were considered but not chosen for $\pi(1)$, together with the $z$ cells in rank 3 of $\sigma$. Since these are two disjoint sets, the number of candidate cells for $\pi(2)$ is $2z$, the same as the number of candidates that we had for $\pi(1)$. The set of cells that were considered but not chosen for $\pi(1)$ are denoted by the set-theoretic difference $U_{1,2}(\sigma)\setminus\pi(1)$. Taking the union of $U_{1,2}(\sigma)\setminus\pi(1)$ with the set $\sigma(3)$, we get that the set of candidate cells to occupy rank 2 of $\pi$ can be denoted by $U_{1,3}(\sigma)\setminus\pi(1)$. 

\emph{Remark:} In the coding of $m_2$, we can in fact use a WOM code with a shorter block length, since the cells in $\pi(1)$ do not need to take any part in the WOM code. This slightly improves the rate and computation complexity of the coding scheme. However, this improvement does not affect the asymptotic analysis we make in this paper. Therefore, for the ease of presentation, we did not use this improvement.

%We now see how WOM codes can help for the construction of RM rewriting codes. Think of $m_1$ as some "part" of the message, say the first few bits of its binary representation. Then the encoder can think of $U_{1,2}(\sigma)$ as a subset of $[n]$, and think of its characteristic vector $\theta_{n}(U_{1,2}(\sigma))$ as the \emph{state} vector $\bfs_1$ of a write-once memory. In this view, an $(n,K_1,2/q,1/q)$ constant-weight strong WOM code $D_W$ can be used for representing $m_1$ by the set $\pi(1)$. First, notice that the weight of $\bfs_1=\theta_{n}(U_{1,2}(\sigma))$ is $2z=(2/q)n$, as required by the WOM code, since $U_{1,2}(\sigma)$ has $2z$ members. Second, the encoder wishes to find a subset $\pi(1)\subset U_{1,2}(\sigma)$ that represents $m_1$. In terms of the characteristic vector, that means that the encoder wishes to find a vector $\bfx_1$ of weight $z=(1/q)n$, such that $\bfx_1\le\bfs_1$, and $\bfx_1$ represents $m_1$. Since we let $m_1$ be represented by a WOM code, Definition \ref{def:cwswom} tells us that such a vector $\bfx_1$ always exists. Furthermore, if the WOM code has an efficient encoding algorithm $E_W$, then we can find $\bfx_1$ fast. Once we find $\bfx_1$, we let $\pi(1)$ be the set with characteristic vector $\bfx_1$, that is $\pi(1)=\theta_{n}^{-1}(\bfx_1)$. When the decoder will read the permutation $\pi$, she could observe the characteristic vector $\theta_{n}(\pi(1))$ and find $m_1$ according to the decoding map of the WOM code.

We now apply the same idea to the rest of the sets of $\pi$, iteratively. On each iteration $i$ from 1 to $q-2$, the set $\pi(i)$ must be a subset of $U_{1,i+1}(\sigma)$, to keep the cost at no more than 1. The sets $\mathset{\pi(1),\dots,\pi(i-1)}$ were already determined in previous iterations, and thus their members cannot belong to $\pi(i)$. The set $U_{1,i-1}(\pi)$ contains the members of those sets (where $U_{1,0}(\pi)$ is the empty set). So we can say that the set $\pi(i)$ must be a subset of $U_{1,i+1}(\sigma)\setminus U_{1,i-1}(\pi)$. We let the state vector of the WOM code to be $\bfs_i=\theta_n(U_{1,i+1}(\sigma)\setminus U_{1,i-1}(\pi))$, and then use the WOM encoder $E_{W}(m_i,\bfs_i)$ to find an appropriate vector $\bfx_i\le\bfs_i$ that represents $m_i$. We then assign $\pi(i)=\theta_n^{-1}(\bfx_i)$, such that $\pi(i)$ represents $m_i$. 

If we use a capacity achieving family of constant-weight strong WOM codes, we store close to $w_sH(w_x/w_s)=2(1/q)H(\frac{1}{2})=2/q$ bits per cell on each rank. Therefore, each of the $q-2$ message parts $m_1,\dots,m_{q-2}$ can store close to $2/q$ bits per cell. So the RM rewriting code can store a total of $2(q-2)/q$ bits per cell, approaching the upper bound of 2 bits per cell (Corollary~\ref{cor:limit_set}) when $q$ grows.
The last message part, $m_{q-1}$, is encoded with the same code $h_M$ that we used in the previous subsection for $q=3$. The amount of information stored in the message $m_{q-1}$ does not affect the asymptotic rate analysis, but is still beneficial.

%To complete the encoding function, we are left with determining the sets $\mathset{\pi(q-1),\dots,\pi(q)}$. These sets can in fact take any permutation of the multiset $M=\mathset{(q-1)^z,(q-2)^z,\dots,q^z}$ without taking the cost of rewrite above $1$. That is because for any cell $j\in U_{q-1,q}(\pi)$, we have $\sigma^{-1}(j)-\pi^{-1}(j)\le q-(q-1)=1$. For that reason, we use an enumerative code for those ranks. Let $h_M:[|\mathfrak{S}_M|]\to\mathfrak{S}_M$ be a bijective mapping from indices to multiset pemutations, and let $h_M^{-1}$ be the inverse map. There are several efficient implementations of such maps, see for example \cite{MilVas00}. The part of the message that will be stored in the ranks $[q-1:q]$ is denoted by $m_{q-1}$, and the sequence of sets $(\pi(q-1),\dots,\pi(q))$ is denoted by $\pi_{[q-1:q]}$, where $\pi_{[q-1:q]}$ is in fact a permutation of the multiset $M$. Then the last step of the encoding algorithm is to assign $\pi_{[q-1:q]}= h_M(m_{q-1})$.

To decode a message vector $\bfm=(m_1,m_2,\dots,m_{q-1})$ from the stored permutation $\pi$, we can just decode each of the $q-1$ message parts separately. For each rank $i\in[q-2]$, the decoder finds the vector $\bfx_i=\theta_n(\pi(i))$, and then the message part $m_i$ is calculated by the WOM decoder, $m_i\Leftarrow D_W(\bfx_i)$. The message part $m_{q-1}$ is found by the decoder of the enumerative code, $m_{q-1}= h_M^{-1}(\pi_{[q-1:q]})$. 

%The \textbf{encoding} algorithm $E_R$ receives a message $\bfm=(m_1,m_2,\dots,m_{q-1})\in[K_W]^{q-2}\times[K_M]$ and a state permutation $\sigma\in \mathfrak{S}_{q,z}$, and returns a permutation $\pi$ in $B_{q,z,1}(\sigma)\cap D^{-1}_R(\bfm)$ to store in the memory. It is constructed as follows:
%\begin{algorithmic}[1]
%\FOR {$i=1$ to $q-2$}
%\STATE $\bfs_i\Leftarrow\theta_n(U_{1,i+1}(\sigma)\setminus U_{1,i-1}(\pi))$
%\STATE $\bfx_i\Leftarrow E_{W}(m_i,\bfs_i)$
%\STATE $\pi(i)\Leftarrow\theta_n^{-1}(\bfx_i)$
%\ENDFOR 
%\STATE $\pi_{[q-1:q]}\Leftarrow h_M(m_{q-1})$
%\end{algorithmic}
%
%The \textbf{decoding} algorithm $D_R$ receives the stored permutation $\pi\in \mathfrak{S}_{q,z}$, and returns the stored message $\bfm=(m_1,m_2,\dots,m_{q-1})\in[K_W]^{q-2}\times[K_M]$. It is constructed as follows:
%\begin{algorithmic}[1]
%\FOR {$i=1$ to $q-2$} 
%\STATE $\bfx_i\Leftarrow\theta_n(\pi(i))$
%\STATE $m_i\Leftarrow D_W(\bfx_i)$
%\ENDFOR 
%\STATE $m_{q-1}\Leftarrow h_M^{-1}(\pi_{[q-1:q]})$
%\end{algorithmic}

\subsection{Generalizing the Cost Constraint $r$}
\label{ss:gen_r}

We note first that if $r$ is larger than $q-2$, the coding problem is trivial.
When the cost constraint $r$ is between 1 and $q-2$, the top $r+1$ cells of $\pi$ can be occupied by \emph{any} cell, since the magnitude of a drop from a rank at most $q$ to a rank at least $q-r-1$, is at most $r$ ranks. Therefore, we let the top $r+1$ ranks of $\pi$ represents a single message part, named $m_{q-r-1}$. The message part $m_{q-r-1}$ is mapped into the arraignment of the sequence of sets $(\pi(q-r),\pi(q-r+1),\dots,\pi(q))$ by a generalization of the bijection $h_M$, defined by generalizing the multiset $M$ into $M=\mathset{(q-r)^z,(q-r+1)^z,\dots,q^z}$. The efficient coding scheme described in~\cite{MilVas00} for $h_M$ and $h_M^{-1}$ is suitable for any multiset $M$.

\begin{figure}[t]
\centering
\includegraphics[scale=0.5]{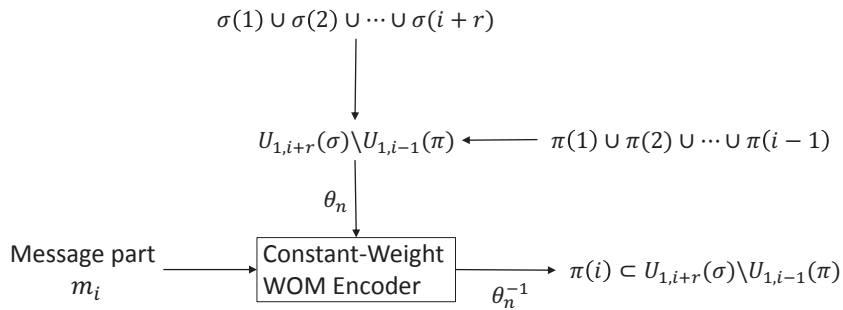}
\caption{Iteration $i$ of the encoding algorithm, where $1\le i\le q-r-1$.}
\label{fig:encoding}
\end{figure}

The rest of the message is divided into $q-r-1$ parts, $m_1$ to $m_{q-r-1}$, and their codes also need to generalized. The generalization of these coding scheme is also quite natural. First, consider the code for the message part $m_1$. When the cost constraint $r$ is larger than 1, more cells are allowed to take rank 1 in $\pi$. Specifically, a cell whose rank in $\sigma$ is at most $r+1$ and its rank in $\pi$ is 1, drops by at most $r$ ranks. Such drop does not cause the rewriting cost to exceed $r$. So the set of candidate cells  for $\pi(1)$ in this case can be taken to be $U_{1,r+1}$. In the same way, for each $i$ in $[1:q-r-1]$, the set of candidate cells for $\pi(i)$ is $U_{1,i+r}(\sigma)\setminus U_{1,i-1}(\pi)$. The parameter $w_s$ of the ingredient WOM is correspondingly generalized to $w_s=(r+1)/q$.
This generalized algorithm is shown in Figure \ref{fig:encoding}. 
We present now a formal description of the construction.

\begin{construction}
\label{con:rmrcs}
{\bf (A RM rewriting code from a constant-weight strong WOM code)} Let $K_W,q,r,z$ be positive integers, let $n=qz$ and let $(E_W,D_{W})$ be an $(n,K_W,(r+1)/q,1/q)$ constant-weight strong WOM coding scheme. Define the multiset $M=\mathset{(q-r)^z,(q-r+1)^z,\dots,q^z}$ and let $K_M=|\mathfrak{S}_M|$ and $K_R=K_W^{q-r-1}\cdot K_M$. The codebook $\cC$ is defined to be the entire set $\mathfrak{S}_{q,z}$. A $(q,z,K_R,r)$ RM rewrite coding scheme $\mathset{E_R,D_R}$ is constructed as follows:

The \textbf{encoding} algorithm $E_R$ receives a message $\bfm=(m_1,m_2,\dots,m_{q-r})\in[K_W]^{q-r-1}\times[K_M]$ and a state permutation $\sigma\in \mathfrak{S}_{q,z}$, and returns a permutation $\pi$ in $B_{q,z,r}(\sigma)\cap D^{-1}_R(\bfm)$ to store in the memory. It is constructed as follows:
\begin{algorithmic}[1]
\FOR {$i=1$ to $q-r-1$}
\STATE $\bfs_i\Leftarrow\theta_n(U_{1,i+r}(\sigma)\setminus U_{1,i-1}(\pi))$
\STATE $\bfx_i\Leftarrow E_{W}(m_i,\bfs_i)$
\STATE $\pi(i)\Leftarrow\theta_n^{-1}(\bfx_i)$
\ENDFOR 
\STATE $\pi_{[q-r:q]}\Leftarrow h_M(m_{q-r})$
\end{algorithmic}

The \textbf{decoding} algorithm $D_R$ receives the stored permutation $\pi\in \mathfrak{S}_{q,z}$, and returns the stored message $\bfm=(m_1,m_2,\dots,m_{q-r})\in[K_W]^{q-r-1}\times[K_M]$. It is constructed as follows:
\begin{algorithmic}[1]
\FOR {$i=1$ to $q-r-1$} 
\STATE $\bfx_i\Leftarrow\theta_n(\pi(i))$
\STATE $m_i\Leftarrow D_W(\bfx_i)$
\ENDFOR 
\STATE $m_{q-r}\Leftarrow h_M^{-1}(\pi_{[q-r:q]})$
\end{algorithmic}
\end{construction}

\begin{theorem}
\label{th:rmrcs}
Let $\mathset{E_W,D_W}$ be a member of an efficient capacity-achieving family of constant-weight strong WOM coding schemes. Then the family of RM rewrite coding schemes in Construction \ref{con:rmrcs} is efficient and capacity-achieving.
\end{theorem}

\begin{IEEEproof}
The decoded message is equal to the encoded message by the property of the WOM code in Definition \ref{def:cwswom}. By the explanation above the construction, it is clear that the cost is bounded by $r$, and therefore $\mathset{E_R,D_R}$ is a RM rewrite coding scheme. We will first show that $\mathset{E_R,D_R}$ is capacity achieving, and then show that it is efficient. Let $R_R=(1/n)\log K_R$ be the rate of a RM rewriting code. To show that $\mathset{E_R,D_R}$ is capacity achieving, we need to show that for any $\epsilon_R>0$, $R_R>C_R-\epsilon_R$, for some $q$ and $z$. 

Since $\mathset{E_W,D_W}$ is capacity achieving, $R_W>C_W-\epsilon_W$ for any $\epsilon_W>0$ and large enough $n$. Remember that $C_W=w_sH(w_x/w_s)$. In $\mathset{E_R,D_R}$ we use $w_s=(r+1)/q$ and $w_x=1/q$, and so $C_W=\frac{r+1}{q}H\left(\frac{1}{r+1}\right)$. We have 
\begin{align}
R_R&=(1/n)\log K_R\nonumber\\
&=(1/n)\log( K_M\cdot K_W^{q-r-1})\nonumber\\
&>(q-r-1)(1/n)\log K_W\nonumber\\
&>(q-r-1)(C_W-\epsilon_W)\label{eq:rate_wom}\\
&=(q-r-1)\left(\frac{r+1}{q}H\left(\frac{1}{r+1}\right)-\epsilon_W\right)\nonumber\\
&=\frac{q-r-1}{q}(C_R-q\epsilon_W)\nonumber\\
&=(C_R-q\epsilon_W)(1-(r+1)/q)\nonumber\\
&>C_R-(r+1)^2/q-q\epsilon_W\nonumber
\end{align}

The idea is to take $q=\lfloor(r+1)/\sqrt{\epsilon_W}\rfloor$ and $\epsilon_R=3(r+1)\sqrt{\epsilon_W}$, and get that 
\begin{equation*}
\label{eq:rate_strong}
R_R>C_R-\frac{(r+1)^2}{\lfloor(r+1)/\sqrt{\epsilon_W}\rfloor}-\lfloor(r+1)/\sqrt{\epsilon_W}\rfloor\epsilon_W>C_R-2(r+1)\sqrt{\epsilon_W}-(r+1)\sqrt{\epsilon_W}=C_R-\epsilon_R.
\end{equation*}
Formally, we say: for any $\epsilon_R>0$ and integer $r$, we set $\epsilon_W=\frac{\epsilon_R^2}{9(r+1)^2}$ and $q=\lfloor(r+1)/\sqrt{\epsilon_W}\rfloor$. Now if $z$ is large enough then $n=qz$ is also large enough so that $R_W>C_W-\epsilon_W$, and then Equation \ref{eq:rate_wom} holds and we have $R_R>C_R-\epsilon_R$, proving that the construction is capacity achieving. Note that the family of coding schemes has a constant value of $q$ and a growing value of $z$, as permitted by Definition \ref{def:capacity_achieving} of capacity-achieving code families.

Next we show that $\mathset{E_R,D_R}$ is efficient. If the scheme $(h_M,h^{-1}_M)$ is implemented as described in \cite{MilVas00}, then the time complexity of $h_M$ and $h_M^{-1}$ is polynomial in $n$. In addition, we assumed that $E_W$ and $D_W$ run in polynimial time in $n$. So since $h_M$ and $h_M^{-1}$ are executed only once in $E_R$ and $D_R$, and $E_W$ and $D_W$ are executed less than $q$ times in $E_R$ and $D_R$, where $q<n$, we get that the time complexity of $E_R$ and $D_R$ is polynomial in $n$.
\end{IEEEproof}

%\emph{Remark:} Since the proof of Theorem \ref{th:rmrcs} is asymptotic, the $\log K_M$ bits stored in the top $r+1$ ranks were not used in the rate analysis. Nonetheless, they could contribute to the performance of practical systems.

\subsection{How to Use Weak WOM Schemes}
\label{ss:weak}

As mentioned earlier, we are not familiar with a family of efficient capacity-achieving constant-weight strong WOM coding schemes. Nonetheless, it turns out that we can construct efficient capacity-achieving WOM coding schemes that meet a slightly weaker definition, and use them to construct capacity-achieving RM rewriting codes. In this subsection we will define a weak notion of constant-weight WOM codes, and describe an associated RM rewriting coding scheme. In Sections \ref{sec:polar} and \ref{sec:hash} we will present yet weaker definition of WOM codes, together with constructions of appropriate WOM schemes and associated RM rewriting schemes.

In the weak definition of WOM codes, each codeword is a \emph{pair}, composed of a constant-weight binary vector $\bfx$ and an index integer $m_a$. Meanwhile, the state is still a single vector $\bfs$, and the vector $\bfx$ in the codeword is required to be smaller than the state vector. We say that these codes are weaker since there is no restriction on the index integer in the codeword. This allows the encoder to communicate some information to the decoder without restrictions.

\begin{definition} 
\label{def:cwwwom}
{\bf (Constant-weight weak WOM codes)} Let $K_W,K_a,n$ be positive integers and let $w_s$ be a real number in $[0,1]$ and $w_x$ be a real number in $[0,w_s]$. A surjective function $D_W: J_{w_x}(n)\times [K_a]\to[K_W]$ is an $(n,K_W,K_a,w_s,w_x)$ \textbf{constant-weight weak WOM code} if for each message $m\in[K_W]$ and state vector $\bfs\in J_{w_s}(n)$, there exists a pair $(\bfx,m_a)$ in the subset $D_W^{-1}(m)\subseteq J_{w_x}(n)\times[K_a]$ such that $\bfx\le\bfs$. The rate of a constant-weight weak WOM code is defined to be $R_W=(1/n)\log(K_W/K_a)$.
\end{definition}

If $K_a=1$, the code is in fact a constant-weight strong WOM code. We will only be interested in the case in which $K_W\gg K_a$.
Since $R_W$ is a decreasing function of $K_a$, it follows that the capacity of constant-weight weak WOM code is also $C_W=w_sH(w_x/w_s)$. Consider now the encoder $E_R$ of a $(q,z,K_R,r)$ RM rewriting code $D_R$ with a codebook $\cC$. For a message $m\in[K_R]$ and a state permutation $\sigma\in \cC$, the encoder needs to find a permutation $\pi$ in the intersection $B_{q,z,r}(\sigma)\cap D_R^{-1}(m)$. As before, we let the encoder determine the sets $\pi(1),\pi(2),\dots,\pi(q-r-1)$ sequentially, such that each set $\pi(i)$ represents a message part $m_i$. If we were to use the previous encoding algorithm (in Construction \ref{con:rmrcs}) with a weak WOM code, the WOM encoding would find a pair $(\bfx_i,m_{a,i})$, and we could store the vector $\bfx_i$ by the set $\pi(i)$. However, we would not have a way to store the index $m_{a,i}$ that is also required for the decoding. To solve this, we will \emph{add} some cells that will serve for the sole purpose of storing the index $m_{a,i}$. 

Since we use the WOM code $q-r-1$ times, once for each rank $i\in[q-r-1]$, it follows that we need to add $q-r-1$ different sets of cells. The added sets will take part in a larger permutation, such that the code will still meet Definition \ref{def:rmrc} of RM rewriting codes. To achieve that property, we let each added set of cells to represent a permutation. That way the number of cells in each rank is constant, and a concatenation (in the sense of sting concatenation) of those permutations together results in a larger permutation. To keep the cost of rewriting bounded by $r$, we let each added set to represent a permutation with $r+1$ ranks. That way each added set could be rewritten arbitrarily with a cost no greater than $r$. We also let the number of cells in each rank in those added sets to be equal, in order to maximize the amount of stored information. Denote the number of cells in each rank in each of the added set as $a$. Since each added set needs to store an index from the set $[K_a]$ with $r+1$ ranks, it follows that $a$ must satisfy the inequality $|\mathfrak{S}_{r+1,a}|\ge K_a$. So to be economical with our resources, we set $a$ to be the smallest integer that satisfies this inequality. We denote each of these additional permutations as $\pi_{a,i}\in\mathfrak{S}_{r+1,a}$. The main permutation is denoted by $\pi_W$, and the number of cells in each rank in $\pi_W$ is denoted by $z_W$. The permutation $\pi$ will be a string concatenation of the main permutation together with the $q-r-1$ added permutations. Note that this way the number of cells in each rank is not equal (there are more cells in the lowest $r+1$ ranks). This is actually not a problem, but it will be cleaner to present the construction if we add yet another permutation that ``balances" the code. Specifically, we let $\pi_b$ be a permutation of the multiset $\mathset{(r+2)^{(q-r-1)a},(r+3)^{(q-r-1)a},\dots,q^{(q-r-1)a}}$ and let $\pi^{-1}$ be the string concatenation $(\pi^{-1}_{a,1},\dots,\pi^{-1}_{a,q-r-1},\pi^{-1}_b,\pi^{-1}_W)$. This way in each rank there are exactly $z_W+(q-r-1)a$ cells. We denote $z=z_W+(q-r-1)a$, and then we get that $\pi$ is a member of $\mathfrak{S}_{q,z}$.

On each iteration $i$ from 1 to $q-r-1$ we use a constant-weight weak WOM code. The vectors $\bfs_i$ and $\bfx_i$ of the WOM code are now corresponding only to the main part of the permutation, and we denote their length by $n_W=qz_W$. We assign the state vector to be $\bfs_i=\theta_{n_W}(U_{1,i+r}(\sigma_W)\setminus U_{1,i-1}(\pi_W))$, where $\sigma_W$ and $\pi_W$ are the main parts of $\sigma$ and $\pi$, accordingly. Note that $U_{1,i+r}(\sigma_W)$ and $U_{1,i-1}(\pi_W)$ are subsets of $[n_W]$ and that the characteristic vector $\theta_{n_W}$ is taken according to $n_W$ as well. The message part $m_i$ and the state vector $\bfs_i$ are used by the encoder $E_W$ of an $(n_W,K_W,K_b,(r+1)/q,1/q)$ constant-weight weak WOM code $D_W$. The result of the encoding is the pair $(\bfx_i,m_{a,i})= E_W(m_i,\bfs_i)$. The vector $\bfx_i$ is stored on the main part of $\pi$, by assigning $\pi_W(i)=\theta_{n_W}^{-1}(\bfx_i)$. The additional index $m_{a,i}$ is stored on the additional cells corresponding to rank $i$. Using an enumerative code $h_{r+1,a}:[|\mathfrak{S}_{r+1,a}|]\to \mathfrak{S}_{r+1,a}$, we assign $\pi_{a,i}= h_{r+1,a}(m_{a,i})$. After the lowest $q-r-1$ ranks of $\pi_W$ are determined, we determine the highest $r+1$ ranks by setting $\pi_{W,[q-r,q]}= h_M(m_{q-r})$ where $M=\mathset{(q-r)^{z_W},(q-r+1)^{z_W},\dots,q^{z_W}}$.
Finally, the permutation $\pi_b$ can be set arbitrarily, say, to $\sigma_b$.

The decoding is performed in accordance with the encoding. For each rank $i\in[q-r-1]$, we first find $\bfx_i=\theta_{n_W}(\pi_W(i))$ and $m_{a,i}= h^{-1}_{r+1,a}(\pi_{a,i})$, and then assign $m_i= D_W(\bfx_i,m_{a,i})$. Finally, we assign $m_{q-r}= h_M^{-1}(\pi_{W,[q-r:q]})$.

\begin{construction}
\label{con:rmrcw}
{\bf (A RM rewriting code from a constant-weight weak WOM code)} Let $K_W,K_a,q,r$ and $z_W$ be positive integers, and let $n_W=qz_W$. Let $D_W$ be an $(n_W,K_W,K_a,(r+1)/q,1/q)$ constant-weight weak WOM code with encoding algorithm $E_W$, and let $a$ be the smallest integer for which $|\mathfrak{S}_{r+1,a}|\ge K_a$. Define the multiset $M=\mathset{(q-r)^{z_W},(q-r+1)^{z_W},\dots,q^{z_W}}$ and let $K_M=|\mathfrak{S}_M|$ and $K=K_M\cdot K_W^{q-r-1}$. 

Let $z=z_W+(q-r-1)a$ and $n=qz$. Define a codebook $\cC\subset \mathfrak{S}_{q,z}$ as a set of permutations $\pi\in\cC$ in which $\pi^{-1}$ is a string concatenation $(\pi^{-1}_W,\pi^{-1}_{a,1},\dots,\pi^{-1}_{a,q-r-1},\pi^{-1}_b)$ such that the following conditions hold:
\begin{enumerate}
\item $\pi_W\in\mathfrak{S}_{q,z_W}$.
\item For each rank $i\in[q-r-1]$, $\pi_{a,i}\in\mathfrak{S}_{r+1,a}$.
\item  $\pi_{b}$ is a permutation of the multiset $\mathset{(r+2)^{(q-r-1)a},(r+3)^{(q-r-1)a},\dots,q^{(q-r-1)a}}$. 
\end{enumerate}
A $(q,z,K_R,r)$ RM rewrite coding scheme $\mathset{E_R,D_R}$ is constructed as follows:

The \textbf{encoding} function $E_R$ receives a message $\bfm=(m_1,m_2,\dots,m_{q-r})\in[K_W]^{q-r-1}\times[K_M]$ and a state permutation $\sigma\in \cC$, and finds a permutation $\pi$ in $B_{q,z,r}(\sigma)\cap D^{-1}_R(\bfm)$ to store in the memory. It is constructed as follows:
\begin{algorithmic}[1]
\FOR {$i=1$ to $q-r-1$}
\STATE $\bfs_i\Leftarrow\theta_{n_W}(U_{1,i+r}(\sigma_W)\setminus U_{1,i-1}(\pi_W))$
\STATE $(\bfx_i,m_{a,i})\Leftarrow E_W(m_i,\bfs_i)$
\STATE $\pi_W(i)\Leftarrow\theta_{n_W}^{-1}(\bfx_i)$
\STATE $\pi_{a,i}\Leftarrow h_{r+1,a}(m_{a,i})$
\ENDFOR 
\STATE $\pi_{W,[q-r:q]}\Leftarrow h_M(m_{q-r})$
\STATE $\pi_b\Leftarrow\sigma_b$
\end{algorithmic}

The \textbf{decoding} function $D_R$ receives the stored permutation $\pi\in \mathcal{C}$, and finds the stored message $\bfm=(m_1,m_2,\dots,m_{q-r})\in[K_W]^{q-r-1}\times[K_M]$. It is constructed as follows:
\begin{algorithmic}[1]
\FOR {$i=1$ to $q-r-1$} 
\STATE $\bfx_i\Leftarrow\theta_{n_W}(\pi_W(i))$
\STATE $m_{a,i}\Leftarrow h^{-1}_{r+1,a}(\pi_{a,i})$
\STATE $m_i\Leftarrow D_W(\bfx_i,m_{a,i})$
\ENDFOR 
\STATE $m_{q-r}\Leftarrow h_M^{-1}(\pi_{W,[q-r:q]})$
\end{algorithmic}
\end{construction}

\emph{Remark:} To be more economical with our resources, we could use the added sets ``on top of each other", such that the $r+1$ lowest ranks store one added set, the next $r+1$ ranks store another added set, and so on. To ease the presentation, we did not describe the construction this way, since the asymptotic performance is not affected. However, such a method could increase the performance of practical systems. 

\begin{theorem}
\label{th:rmrcw}
Let $\mathset{E_W,D_W}$ be a member of an efficient capacity-achieving family of constant-weight weak WOM coding schemes. Then the family of RM rewrite coding schemes in Construction \ref{con:rmrcw} is efficient and capacity-achieving.
\end{theorem}

The proof of Theorem \ref{th:rmrcw} is similar to that of Theorem \ref{th:rmrcs} and is brought in Appendix \ref{app:rmrcw}.

\section{Constant-Weight Polar WOM Codes}
\label{sec:polar}

In this section we consider the use of polar WOM schemes \cite{BurStr13} for the construction of constant-weight weak WOM schemes. Polar WOM codes do not have a constant weight, and thus require a modification in order to be used in Construction \ref{con:rmrcw} of RM rewriting codes. The modification we propose in this section is exploiting the fact that while polar WOM codes do not have a constant weight, their weight is still \emph{concentrated} around a constant value. This section is composed of two subsections. In the first, we show a general method to construct constant-weight weak WOM codes from WOM codes with concentrated weight. The second subsection describes the construction of polar WOM schemes of Burshtein and Strugatski \cite{BurStr13}, and explains how they could be used as concentrated-weight WOM schemes.

\subsection{Constant-Weight Weak WOM Schemes from Concentrated-Weight Strong Schemes}

We first introduce additional notation. Label the weight of a vector $\bfx$ by $w_H(\bfx)$. For $\delta>0$, let $J_{w_x}(n,\delta)$ be the set of all $n$-bit vectors $\bfx$ such that $|w_x-w_H(\bfx)/n|\le \delta $.

\begin{definition} 
\label{def:cwwom}
{\bf (Concentrated-weight WOM codes)} Let $K_C$ and $n$ be positive integers and let $w_s$ be in $[0,1]$, $w_x$ be in $[0,w_s]$ and $\delta$ in $[0,1]$. A surjective function $D_C: J_{w_x}(n,\delta)\to[K_C]$ is an $(n,K_C,w_s,w_x,\delta)$ \textbf{concentrated-weight  WOM code} if for each message $m\in[K_C]$ and state vector $\bfs\in J_{w_s}(n)$, there exists a vector $\bfx\le\bfs$ in the subset $D_C^{-1}(m)\subseteq J_{w_x}(n,\delta)$.
\end{definition}

From Theorem 1 in \cite{Hee85} and Proposition \ref{prop:wom_capacity} we get that the capacity of concentrated-weight WOM codes in $C_W=w_sH(w_x/w_s)$. We define the notion of efficient capacity-achieving family of concentrated-weight WOM coding schemes accordingly. For the construction of constant-weight weak WOM codes from concentrated-weight WOM codes, we will use another type of enumerative coding schemes. For an integer $n$ and $\delta$ in $[0,1/2]$, let $J_{\le \delta}(n)$ be the set of all $n$-bit vectors of weight at most $\delta n$, and define some bijective function $h_{\le\delta}:\left[\sum_{j=1}^{\lfloor\delta n\rfloor}\binom{n}{j}\right]\to J_{\le\delta}(n)$ with an inverse function $h_{\le\delta}^{-1}$. The enumeration scheme $(h_{\le\delta},h_{\le \delta}^{-1})$ can be implemented with computational complexity polynomial in $n$ by methods such as \cite[pp. 27-30]{Bec64}, \cite{Ram90,TiaVaiSlo09}.

We will now describe a construction of a constant-weight weak WOM coding scheme from a concentrated-weight WOM coding scheme. We start with the encoder $E_W$ of the constant-weight weak WOM codes. According to Definition \ref{def:cwwwom}, given a message $m\in[K_W]$ and a state $\bfs\in J_{w_s}(n)$, the encoder needs to find a pair $(\bfx,m_a)$ in the set $D^{-1}_W(m)$ such that $\bfx\le\bfs$. We start the encoding by finding the vector $\bfx_C=E_C(m,\bfs)$ by the encoder of an $(n,K_C,w_s,w_x,\delta)$ concentrated-weight WOM code. We know that the weight of $\bfx_C$ is ``$\delta$-close" to $w_x n$, but we need to find a vector with weight exactly $\lfloor w_x n\rfloor$. To do this, the main idea is to "flip" $|\lfloor w_xn\rfloor-w_H(\bfx_C)|$ bits in $\bfx_C$ to get a vector $\bfx\le\bfs$ of weight $\lfloor w_xn\rfloor$, and store the location of the flipped bits in $m_a$. Let $\bfa$ be the $n$-bit vector of the flipped locations, such that $\bfx=\bfx_C\oplus\bfa$ where $\oplus$ is the bitwise XOR operation. It is clear that the weight of $\bfa$ must be $|\lfloor w_xn\rfloor-w_H(\bfx_C)|$. Let $\bfx_C=(x_{C,1},x_{C,2},\dots,x_{C,n})$.
If $w_H(\bfx_C)<w_xn$, we also must have $a_i=0$ wherever $x_{C,i}=1$, since we only want to flip 0's to 1's to increase the weight. In addition, we must have $a_i=0$ wherever $s_i=0$, since in those locations we have $x_{C,i}=0$ and we want to get $x_i\le s_i$. We can summarize those conditions by requiring that $\bfa\le\bfs\oplus\bfx_C$ if $w_H(\bfx_C)<w_xn$. In the case that $w_H(\bfx_C)>w_xn$, we should require that $\bfa\le\bfx_C$, since $a_i$ can be 1 only where $x_{C,i}=1$. In both cases we have the desired properties $\bfx\le\bfs$, $w_H(\bfx)=\lfloor w_xn\rfloor$ and $w_H(\bfa)\le\delta n$.

To complete the encoding, we let $m_a$ be the index of the vector $\bfa$ in an enumeration of the $n$-bit vectors of weight at most $\delta n$. That will minimize the space required to store $\bfa$. Using an enumerative coding scheme, we assign $m_a= h_{\le\delta}^{-1}(\bfa)$. The decoding is now straight forward, and is described in the following formal description of the construction.

\begin{construction}
\label{con:cwwwom}
{\bf (A constant-weight weak WOM code from a concentrated-weight WOM code)} Let $K_C$ and $n$ be positive integers and let $w_s$ be in $[0,1]$, $w_x$ be in $[0,w_s]$ and $\delta$ in $[0,1/2]$. Let $D_C$ be an $(n,K_C,w_s,w_x,\delta)$ concentrated-weight WOM code, and define $K_W=K_C$ and $K_a= \sum_{i=0}^{\lfloor \delta n\rfloor}\binom{n}{i}$.

An $(n,K_W,K_a,w_s,w_x)$ constant-weight weak WOM coding scheme $\mathset{E_W,D_W}$ is defined as follows:

The \textbf{encoding} function $E_W$ receives a message $m\in[K_W]$ and a state vector $\bfs\in J_{w_x}(n)$, and finds a pair $(\bfx,m_a)$ in $D^{-1}_W(m)\subseteq J_{w_x}(n)\times[K_a]$ such that $\bfx\le\bfs$. It is constructed as follows: 
\begin{enumerate}
\item Let $\bfx_C\Leftarrow E_C(\bfs,m)$. 
\item Let $\bfa$ be an arbitrary vector of weight $|\lfloor w_xn\rfloor-w_H(\bfx_C)|$ such that $\bfa\le\bfs\oplus\bfx_C$ if $w_H(\bfx_C)\le w_x n$ and $\bfa\le\bfx_C$ otherwise. 
\item Return the pair $(\bfx,m_a)\Leftarrow(\bfx_C\oplus\bfa, h^{-1}_{\le\delta}(\bfa))$.
\end{enumerate}

The \textbf{decoding} function $D_W$ receives the stored pair $(\bfx,m_a)\in J_{w_x}(n)\times[K_a]$, and finds the stored message $m\in[K_W]$. It is constructed as follows:
\begin{enumerate}
\item Let $\bfa\Leftarrow  h_{\le\delta}(m_a)$.
\item Let $\bfx_C\Leftarrow \bfx\oplus\bfa$. 
\item Return $m\Leftarrow D_C(\bfx_C)$.
\end{enumerate}
\end{construction}

\begin{theorem}
\label{th:cwwom}
Let $\mathset{E_C,D_C}$ be a member of an efficient capacity-achieving family of concentrated-weight WOM coding schemes. Then Construction \ref{con:cwwwom} describes an efficient capacity-achieving family of constant-weight weak WOM coding schemes for a sufficiently small $\delta$.
\end{theorem}

\begin{IEEEproof}
First, since $E_C,D_C,h_{\le\delta}$ and $h^{-1}_{\le\delta}$ can be performed in polynomial time in $n$, it follows directly that $E_W$ and $D_W$ can also be performed in polynomial time in $n$. Next, we show that the family of coding schemes is capacity achieving. For large enough $n$ we have $(1/n)\log K_W>C_W-\epsilon_C$. So
$$R_W=(1/n)\log(K_W/K_a)>C_W-\epsilon_C-H(\delta),$$
since $\log K_a=\log \sum_{i=0}^{\lfloor \delta n\rfloor}\binom{n}{i}\le nH(\delta)$ by Stirling's formula.
Now, given $\epsilon_W>0$, we let $\epsilon_C=\epsilon_W/2$ and $\delta=H^{-1}(\epsilon_W/2)$ such that $\epsilon_C+H(\delta)=\epsilon_W$. So for large enough $n$ we have $R_W>C_W-\epsilon_C-H(\delta)=C_W-\epsilon_W$.
\end{IEEEproof}

\subsection{Polar WOM Codes}

There are two properties of polar WOM coding schemes that do not fit well in our model. First, the scheme requires the presence of common randomness, known both to the encoder and to the decoder. Such an assumption brings some weakness to the construction, but can find some justification in a practical applications such as flash memory devices. For example, the common randomness can be the address of the storage location within the device. Second, the proposed encoding algorithm for polar WOM coding schemes does not always succeed in finding a correct codeword for the encoded message. In particular the algorithm is randomized, and it only guarantees to succeed with high probability, over the algorithm randomness and the common randomness. Nonetheless, for flash memory application, this assumption can be justified by the fact that such failure probability is much smaller than the unreliable nature of the devices. Therefore, some error-correction capability must be included in the construction for such practical implementation, and a failure of the encoding algorithm will not significantly affect the decoding failure rate. More approaches to tackle this issue are described in \cite{BurStr13}.

The construction is based on the method of channel polarization, which was first proposed by Arikan in his seminal paper \cite{Ari09} in the context of channel coding. We describe it here briefly by its application for WOM coding. This application is based on the use of polar coding for lossy source coding, that was proposed by Korada and Urbanke \cite{KorUrb10}.

Let $n$ be a power of 2, and let
$G_2=\left( \begin{array}{ccc}
1 & 0\\
1 & 1\end{array}\right)$ and $G_2^{\otimes \log n}$ be its $\log n$-th Kronecker product. Consider a memoryless channel with a binary-input and transition probability $W(y|x)$. Define a vector $\bfu\in\{0,1\}^n$, and let $\bfx = \bfu G_2^{\otimes \log n}$, where the matrix multiplication is over $\mathbb{F}_2$. The vector $\bfx$ is the input to the channel, and $\bfy$ is the output vector. The main idea of polar coding is to define $n$ sub-channels
\[ W_n^{(i)}(\bfy,\bfu_{[i-1]}|u_i)=P(\bfy,\bfu_{[i-1]}|u_i)=\frac{1}{2^{n-1}}\sum_{\bfu_{[i+1:n]}}P(\bfy|\bfu).\]
For large $n$, each sub-channel is either very reliable or very noisy, and therefore it is said that the channel is polarized. A useful measure for the reliability of a sub-channel $W_n^{(i)}$ is its Bhattacharyya parameter, defined by
\begin{equation}
\label{eq:polarization}
Z(W_n^{(i)})=\sum_{y\in \cY}\sqrt{W_n^{(i)}(y|0)W_n^{(i)}(y|1)}.
\end{equation}

Consider now a write-once memory. Let $\bfs\in\{0,1\}^n$ be the state vector, and let $w_s$ be the fraction of 1's in $\bfs$. In addition, assume that a user wishes to store the message $m\in K_C$ with a codeword $\bfx\in J_{w_x}(n,\delta)$. The following scheme allows a rate arbitrarily close to $C_W$ for $n$ sufficiently large. The construction uses a compression scheme, based on a \emph{test channel}. Let $v$ be a binary  input to the channel, and $(s,g)$ be the output, where $s$ and $g$ are binary variables as well. Denote $x=g\oplus v$. The probability transition function of the channel is given by 
\[ W(s,g|v)=\left\{\begin{array}{l l}
w_s-w_x & \text{if } (s,x)=(1,0),\\
w_x& \text{if } (s,x)=(1,1),\\
1-w_s & \text{if } (s,x)=(0,0),\\
0 & \text{if } (s,x)=(0,1).\\
\end{array}
\right.
\]

The channel is polarized by the sub-channels $W_n^{(i)}$ of Equation \ref{eq:polarization}, and a \emph{frozen set} $F$ is defined by
$$F=\left \{i\in[n]:Z(W_n^{(i)})\ge 1-2\delta_n^2\right\},$$
where $\delta_n=2^{-n^{\beta}}/(2n)$,
for $0<\beta<1/2$. It is easy to show that the capacity of the test channel is $C_T=1-C_W$. It was shown in \cite{KorUrb10} that $|F|=n(C_T+\epsilon_C)=n(1-C_W+\epsilon_C)$,
where $\epsilon_C$ is arbitrarily small for $n$ sufficiently large. Let $\bfg$ be  a common randomness source from an $n$ dimensional uniformly distributed random binary vector. The coding scheme is the following:

\begin{construction}
\label{con:polar}
{\bf (A Polar WOM  code \cite{BurStr13})} Let $n$ be a positive integer and let $w_s$ be in $[0,1]$, $w_x$ be in $[0,w_s]$ and $\delta$ in $[0,1/2]$. Let $\epsilon_C$ be in $[0,1/2]$ such that $K_C=2^{n(C_W-\epsilon_C)}$ is an integer.

The \textbf{encoding} function $E_C$ receives a message $\bfm\in\{0,1\}^{\lceil \log K_C\rceil}$, a state vector $\bfs\in J_{ w_s}(n)$ and the dither vector $\bfg\in\{0,1\}^n$, and returns a vector $\bfx\le\bfs$ in $D^{-1}_C(\bfm)\subseteq J_{w_x}(n,\delta)$ with high probability. It is constructed as follows: 
\begin{enumerate}
\item Assign $y_j=(s_{j},g_{j})$ and $\bfy=(y_1,y_2,\dots,y_n)$. 
\item Define a vector $\bfu\in\{0,1\}^n$ such that $\bfu_{F}=\bfm$.
\item Create a vector $\hat{\bfu}\in\{0,1\}^n$ by compressing the vector $\bfy$ according to the following successive cancellation scheme: For $i=1,2,\dots,n$, let $\hat{u_i}=u_i$ if $i\in F$. Otherwise, let
\[ \hat{u_i}=\left\{\begin{array}{l l}
0 &\quad \text{w.p. } L_n^{(i)}/(L_n^{(i)}+1)\\
1 &\quad \text{w.p. } 1/(L_n^{(i)}+1)\\
\end{array}
\right.,
\]
where w.p. denotes with probability and 
\[ L_n^{(i)}=L_n^{(i)}(\bfy,\hat{\bfu}_{[i-1]})=\frac{W_n^{(i)}(\bfy,\hat{\bfu}_{[i-1]}|u_i=0)}{W_n^{(i)}(\bfy,\hat{\bfu}_{[i-1]}|u_i=1)}.\]
\item Assign $\bfv\Leftarrow\hat{\bfu} G_2^{\otimes \log n}$.
\item Return $\bfx\Leftarrow\bfv\oplus\bfg$.
\end{enumerate}

The \textbf{decoding} function $D_C$ receives the stored vector $\bfx\in J_{w_x}(n,\delta)$ and the dither vector $\bfg\in\{0,1\}^n$, and finds the stored message $\bfm\in\{0,1\}^{\lceil \log K_C\rceil}$. It is constructed as follows: 
\begin{enumerate}
\item Assign $\bfv\Leftarrow \bfx\oplus\bfg$.
\item Assign $\hat{\bfu}\Leftarrow\bfv (G_2^{\otimes \log n})^{-1}$.
\item Return $\bfm\Leftarrow\hat{\bfu}_F$.
\end{enumerate}
\end{construction}

In \cite{BurStr13}, a few slight modifications for this scheme are described, for the sake of the proof. We use the coding scheme $(E_C,D_C)$ of Construction \ref{con:polar} as an $(N,K_C,w_s,w_x,\delta)$ concentrated-weight WOM coding scheme, even though it does not meet the definition precisely.

By the proof of Lemma 1 of \cite{BurStr13}, for $0<\beta<1/2$, the vector $\bfx$ found by the above encoding algorithm is in $D_C^{-1}(\bfm)$ and in $J_{w_x}(n,\delta)$ w.p. at least $1-2^{-n^{\beta}}$ for $n$ sufficiently large. Therefore, the polar WOM scheme of Construction \ref{con:polar} can be used as a practical concentrated-weight WOM coding scheme for the construction of RM rewriting codes by Constructions \ref{con:rmrcw} and \ref{con:cwwwom}. Lemma 1 of \cite{BurStr13} also proves that this scheme is capacity achieving. By the results in \cite{KorUrb10}, the encoding and the decoding complexities are $O(n\log n)$, and therefore the scheme is efficient. This completes our first full description of a RM rewrite coding scheme in this paper, although it does not meet the definitions of Section \ref{sec:mods} precisely. In the next section we describe a construction of efficient capacity-achieving RM rewrite coding schemes that meet the definitions of Section \ref{sec:mods}.

\section{Rank-Modulation Schemes from Hash WOM Schemes}
\label{sec:hash}

The construction in this section is based on a recent construction of WOM codes by Shpilka \cite{Shp12}. This will require an additional modification to Construction \ref{con:rmrcw} of RM rewrite coding schemes.

\subsection{Rank-Modulation Schemes from Concatenated WOM Schemes}

The construction of Shpilka does not meet any of our previous definitions of WOM codes. Therefore, we define yet another type of WOM codes, called "constant-weight concatenated WOM codes". As the name implies, the definition is a string concatenation of constant-weight WOM codes.

\begin{definition} 
\label{def:cwcwom}
{\bf (Constant-weight concatenated WOM codes)} Let $K_W,K_a,n$ and $t$ be positive integers and let $w_s$ be a real number in $[0,1]$ and $w_x$ be a real number in $[0,w_s]$. A surjective function $D_W: (J_{w_x}(n))^t\times [K_a]\to[K_W]$ is an $(n,t,K_W,K_a,w_s,w_x)$ \textbf{constant-weight concatenated WOM code} if for each message $m\in[K_W]$ and state vector $\bfs\in (J_{ w_s}(n))^t$, there exists a pair $(\bfx,m_a)$ in the subset $D_W^{-1}(m)\subseteq (J_{w_x}(n))^t\times[K_b]$ such that $\bfx\le\bfs$.
\end{definition}

Note that the block length of constant-weight concatenated WOM codes is $nt$, and therefore their rate is defined to be $R_W=\frac{1}{nt}\log K_W$.  Since concatenation does not change the code rate, the capacity of constant-weight concatenated WOM codes is $C_W=w_sH(w_x/w_s)$.
We define the notion of coding schemes, capacity achieving and efficient family of schemes accordingly. Next, we use constant-weight concatenated WOM coding schemes to construct RM rewrite coding schemes by a similar concatenation.

\begin{construction}
\label{con:rmrcc}
{\bf (A RM rewriting scheme from a constant-weight concatenated WOM scheme)} Let $K_W,K_a,q,r,t$ and $z_W$ be positive integers, and let $n_W=qz_W$. Let $D_W$ be an $(n_W,t,K_W,K_a,(r+1)/q,1/q)$ constant-weight concatenated WOM code with encoding algorithm $E_W$, and let $a$ be the smallest integer for which $|\mathfrak{S}_{r+1,a}|\ge K_b$. Define the multiset $M=\mathset{(q-r)^{z_W},(q-r+1)^{z_W},\dots,q^{z_W}}$ and let $K_M=|\mathfrak{S}_M|$ and $K_R=K_M\cdot K_W^{q-r-1}$. 

Let $z=tz_W+(q-r-1)a$ and $n=qz$. Define a codebook $\cC\subset \mathfrak{S}_{q,z}$ as a set of permutations $\pi\in\cC$ in which $\pi^{-1}$ is a string concatenation $(\pi^{-1}_{a,1},\dots,\pi^{-1}_{a,q-r-1},\pi^{-1}_b,\pi^{-1}_{x,1},\dots,\pi^{-1}_{x,t})$ such that the following conditions hold:
\begin{enumerate}
\item $\pi_{x,i}\in\mathfrak{S}_{q,z_W}$ for each $i\in[t]$.
\item $\pi_{a,i}\in\mathfrak{S}_{r+1,a}$ for each rank $i\in[q-r-1]$.
\item $\pi_{b}$ is a permutation of the multiset $\mathset{(r+2)^{(q-r-1)a},(r+3)^{(q-r-1)a},\dots,q^{(q-r-1)a}}$. 
\end{enumerate}
Denote the string concatenation $(\pi^{-1}_{x,1},\dots,\pi^{-1}_{x,t})$ by $\pi_W^{-1}$, and denote $\sigma_W$ in the same way.
A $(q,z,K_R,r)$ RM rewrite coding scheme $\mathset{E_R,D_R}$ is constructed as follows:

The \textbf{encoding} function $E_R$ receives a message $\bfm=(m_1,m_2,\dots,m_{q-r})\in[K_W]^{q-r-1}\times[K_M]$ and a state permutation $\sigma\in \cC$, and finds a permutation $\pi$ in $B_{q,z,r}(\sigma)\cap D^{-1}_R(\bfm)$ to store in the memory. It is constructed as follows:
\begin{algorithmic}[1]
\FOR {$i=1$ to $q-r-1$}
\STATE $\bfs_i\Leftarrow\theta_{n_W}(U_{1,i+r}(\sigma_W)\setminus U_{1,i-1}(\pi_W))$
\STATE $(\bfx_i,m_{a,i})\Leftarrow E_W(m_i,\bfs_i)$
\STATE $\pi_W(i)\Leftarrow\theta_{n_W}^{-1}(\bfx_i)$
\STATE $\pi_{a,i}\Leftarrow h_{r+1,a}(m_{a,i})$
\ENDFOR 
\STATE $\pi_{W,[q-r,q]}\Leftarrow h_M(m_{q-r})$
\STATE $\pi_b=\sigma_b$
\end{algorithmic}

The \textbf{decoding} function $D_R$ receives the stored permutation $\pi\in \mathcal{C}$, and finds the stored message $\bfm=(m_1,m_2,\dots,m_{q-r})\in[K_W]^{q-r-1}\times[K_M]$. It is constructed as follows:
\begin{algorithmic}[1]
\FOR {$i=1$ to $q-r-1$} 
\STATE $\bfx_i\Leftarrow\theta_{n_W}(\pi_W(i))$
\STATE $m_{a,i}\Leftarrow h^{-1}_{r+1,a}(\pi_{a,i})$
\STATE $m_i\Leftarrow D_W(\bfx_i,m_{a,i})$
\ENDFOR 
\STATE $m_{q-r}\Leftarrow h_M^{-1}(\pi_{W,[q-r,q]})$
\end{algorithmic}
\end{construction}

Since again concatenation does not affect the rate of the code, the argument of the proof of Theorem \ref{th:rmrcw} gives the following statement:
\begin{theorem}
Let $\mathset{E_W,D_W}$ be a member of an efficient capacity-achieving family of constant-weight concatenated WOM coding schemes. Then the family of RM rewrite coding schemes in Construction \ref{con:rmrcc} is efficient and capacity-achieving.
\end{theorem}

\subsection{Hash WOM Codes}

In \cite{Shp12} Shpilka proposed a construction of efficient capacity-achieving WOM coding scheme. The proposed scheme follows the concatenated structure of Definition \ref{def:cwcwom}, but does not have a constant weight. In this subsection we describe a slightly modified version of the construction of Shpilka, that does exhibit a constant weight. 

To describe the construction, we follow the definitions of Shpilka \cite{Shp12}. The construction is based on a set of hash functions. For positive integers $n,k,l$ and field members $a,b\in\mathbb{F}_{2^n}$, define a map $H_{a,b}^{n,k,l}:\{0,1\}^n\to\{0,1\}^{k-l}$ as $H_{a,b}^{n,k,l}(\bfx)=(ax+b)_{[k-l]}$. This notation means that we compute the affine transformation $ax+b$ in $\mathbb{F}_{2^n}$, represent it as a vector of $n$ bits using the natural map and then keep the first $k-l$ bits of this vector. We represent this family of maps by $\cH^{n,k,l}$, namely
$$\cH^{n,k,l}=\left \{H_{a,b}^{n,k,l}| a,b\in\mathbb{F}_{2^n}\right \}.$$
The family $\cH^{n,k,l}$ contains $2^{2n}$ functions. For an integer $m_a\in[2^{2n}]$, we let $H_{m_a}$ be the $m_a$-th function in $\cH^{n,k,l}$.

\begin{construction}
\label{con:hash}
{\bf (A constant-weight concatenated WOM coding scheme from hash functions)}
Let $\epsilon,\delta$ be in $[0,1/2]$, $w_s$ in $[0,1]$, $w_x$ in $[0,w_s]$ and $c>20$. Let $n=\lceil(c/\epsilon)\log(1/\epsilon)\rceil$, $k=\lfloor n (C_W-2\epsilon /3)\rfloor$, $t_1=\lfloor(1/\epsilon)^{c/12}-1\rfloor$ and $t_2=2^{\frac{4n}{\delta}}$. Finally, Let $t=t_1t_2$, $K_b=2^{k}$ and $K_a=2^{2n}$.
An $(n,t,K_b^t,K_a^{t_2},w_s,w_x)$ constant-weight concatenated WOM code is defined as follows:

The \textbf{encoding} function $E_W$ receives a message matrix $\bfm\in[K_b]^{t_1\times t_2}$, a state matrix of vectors $\bfs\in (J_{w_s}(n))^{t_1\times t_2}$, and returns a pair $(\bfx,\bfm_a)$ in $D^{-1}_W(\bfm)\subseteq (J_{w_x}(n))^{t_1\times t_2}\times[K_a]^{t_2}$ such that for each $(i,j)\in[t_1]\times[t_2]$ we have $\bfx_{i,j}\le\bfs_{i,j}$. It is constructed as follows: For each $j\in[t_2]$, use a brute force search to find an index $m_{a,j}\in[K_a]$ and a vector $\bfx_j=(\bfx_{1,j},\dots,\bfx_{t_1,j})$ such that for all $i\in[t_1]$, the following conditions hold:
\begin{enumerate}
\item $\bfx_{i,j}\le \bfs_{i,j}$.
\item $\bfx_{i,j}\in J_{w_x}(n)$.
\item $H_{m_{a,j}}(\bfx_{i,j})=m_{i,j}$.
\end{enumerate}

The \textbf{decoding} function $D_W$ receives the stored pair $(\bfx,\bfm_a)\in (J_{w_x}(n))^{t_1\times t_2}\times[K_a]$, and returns the stored message $\bfm\in [K_b]^{t_1\times t_2}$. It is constructed as follows: For each pair $(i,j)\in [t_1]\times [t_2]$, assign $m_{i,j}\Leftarrow H_{m_{a,j}}(\bfx_{i,j})$.
\end{construction}

The only conceptual difference between Construction \ref{con:hash} and the construction in \cite{Shp12} is that here we require the vectors $\bfx_{i,j}$ to have a constant weight of $\lfloor w_xn\rfloor$, while the construction in \cite{Shp12} requires the weight of those vectors to be only bounded by $w_xn$. This difference is crucial for the rank-modulation application, but in fact it has almost no effect on the proofs of the properties of the construction.

To prove that the code in Construction \ref{con:hash} is a constant-weight concatenated WOM code, we will need the following lemma from \cite{Shp12}:
\begin{lemma}
\label{lem:23}
\cite[Corollary 2.3]{Shp12}: Let $k',\ell,t_1$ and $n$ be positive integers such that $\ell\le k'\le n$ and $t_1<2^{\ell/4}$. Let $\bfX_1,\dots,\bfX_{t_1}\subseteq\{0,1\}^n$ be sets of size $|\bfX_1|,\dots,|\bfX_{t_1}|\ge 2^{k'}$. Then, for any $\bfm_1,\dots,\bfm_{t_1}\in\{0,1\}^{k'-\ell}$ there exists $H_m\in\cH^{n,k',\ell}$ and $\{\bfx_i\in \bfX_i\}$ such that for all $i\in[t_1]$, $H_m(\bfx_i)=\bfm_i$.
\end{lemma}
Lemma~\ref{lem:23} is proven using the leftover hash lemma \cite[pp. 445]{AroBar09}, \cite{BenBraRob88,ImpLevLub89} and the probabilistic method.

\begin{proposition}
\label{pro:hash_basic}
The code $D_W$ of Construction \ref{con:hash} is an $(n,t,K_b^t,K_a^{t_2},w_s,w_x)$ constant-weight concatenated WOM code.
\end{proposition}

\begin{IEEEproof}
The proof is almost the same as the proof of Lemma 2.4 in \cite{Shp12}, except that here the codewords' weight is constant.
Let $\ell=\lceil\epsilon n/3\rceil$, $k'=k+\ell$ and 
$$\bfX_i=\{\bfx\in\{0,1\}^n|\bfx\le \bfs_i \text{ and } \bfx\in J_{w_x}(n)\}.$$

Since $\bfx\in J_{w_x}(n)$, we have that 
$$|\bfX_i|=\binom{\lfloor w_sn\rfloor}{\lfloor w_xn\rfloor}$$
which by Stirling's formula can be lower bounded by
\begin{align*}
\ge 2^{w_snH(w_x/w_s)-\log(w_sn)}&\ge 2^{nC_W-\log n}\\
&\ge 2^{nC_W-\epsilon n/3}=2^{k'}
\end{align*}

For the last inequality we need $\epsilon n\ge 3\log n$, which follows from
$$\frac{3\log n}{\epsilon n}<\frac{3\log [(2c/\epsilon)\log(1/\epsilon)]}{c\log(1/\epsilon)}<\frac{3\log [(40/\epsilon)\log(1/\epsilon)]}{20\log(1/\epsilon)}<1.$$

Notice also that
$$t_1=\lfloor(1/\epsilon)^{c/12}-1\rfloor<(1/\epsilon)^{c/12}=2^{\frac{1}{4}\frac{\epsilon}{3}\frac{c}{\epsilon}\log (1/\epsilon)}\le 2^{\frac{1}{4}\frac{\epsilon n}{3}}\le 2^{\ell/4}.$$
So all of the conditions of Lemma \ref{lem:23} are met, which implies that the encoding of Construction \ref{con:hash} is always successful, and thus that $D_W$ is a constant-weight concatenated WOM code.
\end{IEEEproof}

\begin{theorem}
Construction \ref{con:hash} describes an efficient capacity-achieving family of concatenated WOM coding schemes.
\end{theorem}

\begin{IEEEproof}
We first show that the family is capacity achieving. We will need the following inequality:
$$\frac{2}{t_1}=\frac{2}{\lfloor(1/\epsilon)^{c/12}-1\rfloor}<4\epsilon^{5/3}<\epsilon/3.$$
Now the rate can be bounded bellow as follows:
\begin{align*}
R_W&=\frac{t\log K_b-t_2\log K_a}{nt}\\
&=\frac{t_1\log K_b-\log K_a}{nt_1}\\
&=\frac{t_1k-2n}{nt_1}\\
&\ge\frac{t_1(C_W-2\epsilon /3)-2}{t_1}\\
&>C_W-2\epsilon/3-\epsilon/3\\
&=C_W-\epsilon,
\end{align*}
and therefore the family is capacity achieving.

To show that the family is efficient, denote the block length of the code as $N=nt$. The encoding time is
$$t_2|\cH^{n,k,\ell}|\cdot\sum_{i=1}^{t_1}|\bfX_i|\le t_2t_12^{3n}<t_2 2^{4n}=t_2^{1+\delta}<N^{1+\delta},$$
and the decoding time is
$$t_2\cdot\text{poly}(kt_1n)=2^{4n/\delta}(2/\epsilon)^{O(c)}<N\cdot 2^{O(n\epsilon)}=N\cdot N^{O(\delta\epsilon)}=N^{1+O(\delta\epsilon)}.$$
This completes the proof of the theorem.
\end{IEEEproof}

\emph{Remark:} Note that $t_2$ is exponential in $1/\epsilon$, and therefore the block length $N$ is exponential in $(1/\epsilon)$. This can be an important disadvantage for these codes. In comparison, it is likely that the block length of polar WOM codes is only polynomial in $(1/\epsilon)$, since a similar results was recently shown in \cite{GolBur13} for the case of polar lossy source codes, on which polar WOM codes are based. 

We also note here that it is possible that the WOM codes of Gabizon and Shaltiel \cite{GabSha12} could be modified for constant weight, to give RM rewriting codes with short block length without the dither and error probability of polar WOM codes.

\section{Conclusions}
\label{sec:conclusions}

In this paper we studied the limits of rank-modulation rewriting codes, and presented two capacity-achieving code constructions. The construction of Section \ref{sec:hash}, based on hash functions, has no possibility of error, but require a long block length that might not be considered practical. On the other hand, the construction of section \ref{sec:polar}, based on polar codes, appears to have a shorter block length, but requires the use of common randomness and exhibit a small probability of error. Important open problems in this area include the rate of convergence of polar WOM codes and the study of error-correcting rewriting codes. Initial results regarding error-correcting polar WOM codes were proposed in \cite{JiaLiEngLanBru13}.

\section{Acknowledgments}
This work was partially supported by the NSF grants ECCS-0801795 and CCF-1217944, NSF CAREER Award CCF-0747415, BSF grant 2010075 and a grant from Intellectual Ventures.

\appendices

\section{}
\label{app:cost}
\begin{IEEEproof}[Proof of Proposition \ref{prop:cost}]
We want to prove that if $\Gamma_{\bfs}(i+1)-\Gamma_{\bfs}(i)\ge1$ for all $i\in[q-1]$, and $\pi$ is in $\mathfrak{S}_M$, then
\begin{equation*}
\alpha(\bfs\to\pi)\le\max_{j\in [n]}\{\sigma_{\bfs}^{-1}(j)-\pi^{-1}(j)\}
\end{equation*}
with equality if $\Gamma_{\bfs}(q)-\Gamma_{\bfs}(1)=q-1$.

The assumption implies that
\begin{equation}
\label{eq:ranks}
\Gamma_{\bfs}(i)\le\Gamma_{\bfs}(q)+i-q
\end{equation}
 for all $i\in[q]$, with equality if $\Gamma_{\bfs}(q)-\Gamma_{\bfs}(1)=q-1$.

%We take $i=q$ as the base case, which holds trivially. Note also that since the cell-state vector $\bfs$ is in $Q_M$, we have $\Gamma_{\bfs}(i)<\Gamma_{\bfs}(i+1)$ for all $i\in[q-1]$. Now for the inductive step we use $\Gamma_{\bfs}(i-1)\le\Gamma_{\bfs}(i)-1$ and the inductive hypothesis and get
%$$\Gamma_{\bfs}(i-1)\le\Gamma_{\bfs}(i)-1\le \Gamma_{\bfs}(q)+i-q -1,$$
 %which proves the induction claim. Note that the condition $\Gamma_{\bfs}(q)-\Gamma_{\bfs}(1)=q-1$ in the statement of the lemma implies that $\Gamma_{\bfs}(i)=\Gamma_{\bfs}(i+1)-1$ for all $i\in [q-1]$, and therefore equality in Equation \ref{eq:ranks}.

Next, define a set $U_{i_1,i_2}(\sigma_{\bfs})$ to be the union of the sets $\mathset{\sigma_{\bfs}(i)}_{i\in[i_1:i_2]}$, and remember that the writing process sets $x_j=s_j$ if $\pi^{-1}(j)=1$, and otherwise 
$$x_j=\max\{s_j,\Gamma_{\bfx}(\pi^{-1}(j)-1)+1\}.$$
Now we claim by induction on $i\in[q]$ that
\begin{equation}
\label{eq:partial_cost}
\Gamma_{\bfx}(i)\le i+\Gamma_{\bfs}(q)-q+\max_{j\in U_{1,i}(\pi)}\{\sigma_{\bfs}^{-1}(j)-\pi^{-1}(j)\}.
\end{equation}

In the base case, $i=1$, and
\begin{align*}
\Gamma_{\bfx}(1)\overset{\text{(a)}}=&\max_{j\in \pi(1)}\{x_j\}\overset{\text{(b)}}=\max_{j\in \pi(1)}\{s_j\}\overset{\text{(c)}}\le\max_{j\in \pi(1)}\{\Gamma_{\bfs}(\sigma_{\bfs}^{-1}(j))\}\overset{\text{(d)}}\le  \max_{j\in \pi(1)}\{\Gamma_{\bfs}(q)-q+\sigma_{\bfs}^{-1}(j)\}\\
\overset{\text{(e)}}=&\Gamma_{\bfs}(q)-q+ \max_{j\in \pi(1)}\{\sigma_{\bfs}^{-1}(j)+(1-\pi^{-1}(j))\}\overset{\text{(f)}}= 1+\Gamma_{\bfs}(q)-q+ \max_{j\in U_{1,i}(\pi)}\{\sigma_{\bfs}^{-1}(j)-\pi^{-1}(j)\}
\end{align*}
Where (a) follows from the definition of $\Gamma_{\bfx}(1)$, (b) follows from the modulation process, (c) follows since $\Gamma_{\bfs}(\sigma_{\bfs}^{-1}(j))=\max_{j'\in \sigma_{\bfs}(\sigma_{\bfs}^{-1}(j))}\{s_{j'}\}$, and therefore $\Gamma_{\bfs}(\sigma_{\bfs}^{-1}(j))\ge s_j$ for all $j\in[n]$ , (d) follows from Equation \ref{eq:ranks}, (e) follows since $j\in \pi(1)$, and therefore $\pi^{-1}(j)=1$, and (f) is just a rewriting of the terms. Note that the condition $\Gamma_{\bfs}(q)-\Gamma_{\bfs}(1)=q-1$ implies that $s_j=\Gamma_{\bfs}(\sigma_{\bfs}^{-1}(j))$ and $\Gamma_{\bfs}(i)=\Gamma_{\bfs}(q)+i-q$, and therefore equality in (c) and (d).

\allowdisplaybreaks
For the inductive step, we have
\begin{align*}
\Gamma_{\bfx}(i)\overset{\text{(a)}}=&\max_{j\in \pi(i)}\{x_j\}\\
\overset{\text{(b)}}=&\max_{j\in \pi(i)}\{\max\{s_j,\Gamma_{\bfx}(i-1)+1\}\}\\
\overset{\text{(c)}}\le& \max\{\max_{j\in \pi(i)}\{s_j\},(i-1)+\Gamma_{\bfs}(q)-q+\max_{j\in U_{1,i-1}(\pi)}\{\sigma_{\bfs}^{-1}(j)-\pi^{-1}(j)\}+1\}\\
\overset{\text{(d)}}\le& \max\{\max_{j\in \pi(i)}\{\Gamma_{\bfs}(\sigma_{\bfs}^{-1}(j))\},i+\Gamma_{\bfs}(q)-q+\max_{j\in U_{1,i-1}(\pi)}\{\sigma_{\bfs}^{-1}(j)-\pi^{-1}(j)\}\}\\
\overset{\text{(e)}}\le& \max\{\max_{j\in \pi(i)}\{\Gamma_{\bfs}(q)-q+\sigma_{\bfs}^{-1}(j)\},i+\Gamma_{\bfs}(q)-q+\max_{j\in U_{1,i-1}(\pi)}\{\sigma_{\bfs}^{-1}(j)-\pi^{-1}(j)\}\}\\
\overset{\text{(f)}}=& \Gamma_{\bfs}(q)-q+\max\{\max_{j\in \pi(i)}\{\sigma_{\bfs}^{-1}(j)+(i-\pi^{-1}(j))\},i+\max_{j\in U_{1,i-1}(\pi)}\{\sigma_{\bfs}^{-1}(j)-\pi^{-1}(j)\}\}\\
\overset{\text{(g)}}=& i+\Gamma_{\bfs}(q)-q+\max\{\max_{j\in \pi(i)}\{\sigma_{\bfs}^{-1}(j)-\pi^{-1}(j)\},\max_{j\in U_{1,i-1}(\pi)}\{\sigma_{\bfs}^{-1}(j)-\pi^{-1}(j)\}\}\\
\overset{\text{(h)}}=& i+\Gamma_{\bfs}(q)-q+\max_{j\in U_{1,i}(\pi)}\{\sigma_{\bfs}^{-1}(j)-\pi^{-1}(j)\}\\
\end{align*}
Where (a) follows from the definition of $\Gamma_{\bfx}(i)$, (b) follows from the modulation process, (c) follows from the induction hypothesis, (d) follows from the definition of $\Gamma_{\bfs}(\sigma_{\bfs}^{-1}(j))$, (e) follows from Equation \ref{eq:ranks}, (f) follows since $\pi^{-1}(j)=i$, and (g) and (h) are just rearrangements of the terms. This completes the proof of the induction claim. As in the base case, we see that if $\Gamma_{\bfs}(q)-\Gamma_{\bfs}(1)=q-1$ then the inequality in Equation \ref{eq:partial_cost} becomes an equality.

Finally, taking $i=q$ in Equation \ref{eq:partial_cost} gives
\[\Gamma_{\bfx}(q)\le q+\Gamma_{\bfs}(q)-q+\max_{j\in U_{1,q}(\pi)}\{\sigma_{\bfs}^{-1}(j)-\pi^{-1}(j)\}=\Gamma_{\bfs}(q)+\max_{j\in [n]}\{\sigma_{\bfs}^{-1}(j)-\pi^{-1}(j)\}\]
with equality if $\Gamma_{\bfs}(q)-\Gamma_{\bfs}(1)=q-1$, which completes the proof of the proposition, since $\alpha(\bfs\to\pi)$ was defined as $\Gamma_{\bfx}(q)-\Gamma_{\bfs}(q)$.
\end{IEEEproof}

\section{}
\label{app:wom_capacity}
\begin{IEEEproof}[Proof of Proposition \ref{prop:wom_capacity}]
The proof follows a similar proof by Heegard \cite{Hee85}, for the case where the codewords' weight is not necessarily constant. Given a state $\bfs$, the number of vectors $\bfx$ of weight $\lfloor w_xn\rfloor$ such that $\bfx\le\bfs$ is $\binom{\lfloor w_s n\rfloor}{\lfloor w_x n\rfloor}$. Since $K_W$ cannot be greater than this number, we have 
$$R_W=(1/n)\log K_W\le (1/n)\log \binom{\lfloor w_s n\rfloor}{\lfloor w_x n\rfloor}\le (1/n)\log 2^{w_s nH(w_x/w_s)}=C_W,$$
where the last inequality follows from Stirling's formula. Therefore, the capacity is at most $C_W$.

The lower bound on the capacity is proven by the probabilistic method. Randomly and uniformly partition $J_{w_x}(n)$ into $K_W$ subsets of equal size, 
$$|D_W^{-1}(m)|=|J_{w_x}(n)|/2^{nR_W}.$$
Fix $m\in[K_W]$ and $\bfs\in J_{w_s}(n)$, and let $\beta(\bfs)$ be the set of vectors $\bfx\in J_{w_x}(n)$ such that $\bfx\le\bfs$. Then
\begin{align*}
P(D_W^{-1}(m)\cap\beta(\bfs)=\emptyset)&=\prod_{i=0}^{|D_W^{-1}(m)|-1}\frac{|J_{w_x}(n)|-|\beta(\bfs)|-i}{|J_{w_x}(n)|-i}\\
&\le\left(\frac{|J_{w_x}(n)|-|\beta(\bfs)|}{|J_{w_x}(n)|}\right)^{|D_W^{-1}(m)|}.
\end{align*}
$|\beta(\bfs)|\ge 2^{n C_W-\log(w_sn)}$, and thus
\begin{align*}
P(D_W^{-1}(m)\cap\beta(\bfs)=\emptyset)&\le(1-|J_{w_x}(n)|^{-1}2^{n C_W-\log(w_sn)})^{|J_{w_x}|2^{-nR_W}}\\
&< e^{-(2^{n(C_W-R_W)-\log(w_sn)})},
\end{align*}
where the last inequality follows from the fact that $(1-x)^y<e^{-xy}$ for $y>0$. If $R_W<C_W$, this probability vanishes for large $n$. In addition,
\begin{align*}
P(\exists m\in[K_W] \text{ and }& \bfs\in J_{w_s}(n) \text{ s.t. } D_W^{-1}(m)\cap\beta(\bfs)=\emptyset)\\
&=P\left(\cup_{m\in[K_W]}\cup_{\bfs\in J_{w_s}(n)}\mathset{D_W^{-1}(m)\cap\beta(\bfs)=\emptyset}\right)\\
&\le\sum_{m\in[K_W]}\sum_{\bfs\in J_{w_s}(n)}P(D_W^{-1}(m)\cap\beta(\bfs)=\emptyset)\\
&\le 2^{n(R_W+H(w_s))}e^{-(2^{n(C_W-R_W)-\log(w_sn)})}
\end{align*}
This means that if $R_W<C_W$ and $n$ is large enough, the probability that the partition is not a constant-weight strong WOM code approaches 0, and therefore there exists such a code, completing the proof.
\end{IEEEproof}

\section{}
\label{app:rmrcw}
\begin{IEEEproof}[Proof of Theorem \ref{th:rmrcw}]
 We will first show that $\mathset{E_R,D_R}$ is capacity achieving, and then show that it is efficient. Let $R_R=(1/n)\log K_R$ be the rate of a RM rewriting code. To show that $\mathset{E_R,D_R}$ is capacity achieving, we need to show that for any $\epsilon_R>0$, $R_R>C_R-\epsilon_R$, for some $q$ and $z$. 

Since $\mathset{E_W,D_W}$ is capacity achieving, $R_W>C_W-\epsilon_W$ for any $\epsilon_W>0$ and large enough $n$. Remember that $C_W=w_sH(w_x/w_s)$. In $\mathset{E_R,D_R}$ we use $w_s=(r+1)/q$ and $w_x=1/q$, and so $C_W=\frac{r+1}{q}H\left(\frac{1}{r+1}\right)$. We will need to use the inequality $\log K_a>a$, which follows from:
$$\log K_a>\log |\mathfrak{S}_{r+1,a-1}|>\log |\mathfrak{S}_{2,a-1}|>2a-2-\log 2a>a$$
Where the last inequality requires $a$ to be at least $6$. In  addition, we will need the inequality $n_W/n>1-q^2\epsilon_W$, which follows form:
\begin{align*}
\frac{n_W}{n}&=\frac{n_W}{n_W+q(q-r-1)a}>\frac{n_W}{n_W+q^2a}>1-\frac{q^2a}{n_W}>1-\frac{q^2\log K_a}{n_W}\\
&=1-q^2\left(\frac{\log K_W}{n_W}-\frac{\log(K_W/K_a)}{n_W}\right)>1-q^2(C_W-(C_W-\epsilon_W))=1-q^2\epsilon_W.
\end{align*}
Now we can bound the rate from below, as follows:
\begin{align}
R_R&=(1/n)\log K_R\nonumber\\
&=(1/n)\log( K_M\cdot K_W^{q-r-1})\nonumber\\
&>(q-r-1)(1/n)\log K_W\nonumber\\
&>(q-r-1)(C_W-\epsilon_W)\label{eq:rate_womw}(n_W/n)\\
&>(q-r-1)\left(\frac{r+1}{q}H\left(\frac{1}{r+1}\right)-\epsilon_W\right)(1-q^2\epsilon_W)\nonumber\\
&=\frac{q-r-1}{q}(C_R-q\epsilon_W)(1-q^2\epsilon_W)\nonumber\\
&=(C_R-q\epsilon_W)(1-(r+1)/q)(1-q^2\epsilon_W)\nonumber\\
&>C_R-C_Rq^2\epsilon_W-C_R(r+1)/q+(C_R(r+1)q\epsilon_W-q\epsilon_W)+(q^3\epsilon^2-(r+1)q^2\epsilon_W^2)\nonumber\\
&>C_R-(r+1)q^2\epsilon_W-(r+1)^2/q\nonumber
\end{align}

The idea is to take $q=\left\lfloor\left(\frac{r+1}{\epsilon_W}\right)^{1/3}\right\rfloor$ and $\epsilon_R=3(r+1)^{2/3}\epsilon_W^{1/3}$ and get that 
\begin{equation*}
\label{eq:rate_weak}
R_R>C_R-(r+1)\left\lfloor\left(\frac{r+1}{\epsilon_W}\right)^{1/3}\right\rfloor^2\epsilon_W-\frac{(r+1)^2}{\left\lfloor\left(\frac{r+1}{\epsilon_W}\right)^{1/3}\right\rfloor}>C_R-(r+1)^{2/3}\epsilon_W^{1/3}-2(r+1)^{2/3}\epsilon_W^{1/3}=C_R-\epsilon_R.
\end{equation*}
So we can say that for any $\epsilon_R>0$ and integer $r$, we set $\epsilon_W=\frac{\epsilon_R^2}{9(r+1)^2}$ and $q=\lfloor(r+1)/\sqrt{\epsilon_W}\rfloor$. Now if $z$ is large enough then $n=qz$ is also large enough so that $R_W>C_W-\epsilon_W$, and then Equation \ref{eq:rate_womw} holds and we have $R_R>C_R-\epsilon_R$.

Finally, we show that $\mathset{E_R,D_R}$ is efficient. If the scheme $(h_M,h^{-1}_M)$ is implemented as described in \cite{MilVas00}, then the time complexity of $h_M$ and $h_M^{-1}$ is polynomial in $n$. In addition, we assumed that $E_W$ and $D_W$ run in polynomial time in $n$. So since $h_M$ and $h_M^{-1}$ are executed only once in $E_R$ and $D_R$, and $E_W$ and $D_W$ are executed less than $q$ times in $E_R$ and $D_R$, where $q<n$, we get that the time complexity of $E_R$ and $D_R$ is polynomial in $n$.
\end{IEEEproof}

\ifCLASSOPTIONcaptionsoff
  \newpage
\fi

\bibliographystyle{IEEEtranS}
\bibliography{allbib}
%

\end{document}